%% file: main.tex
\documentclass[acmsmall]{acmart}

\usepackage{colortbl} 
\usepackage{makecell}
\usepackage{multirow}
\usepackage{graphicx}
\usepackage{float} 
\usepackage{subfigure} 
\usepackage{url}

\AtBeginDocument{%
  }

\setcopyright{acmlicensed}
\copyrightyear{2025}
\acmYear{2025}

\acmConference[CSCW 2025]{The 28th ACM SIGCHI Conference on Computer-Supported Cooperative Work \& Social Computing}{October 18 — 22, 2025}{Bergen, Norway}

\begin{document}

\title[Exploring the Effects of Chatbot Anthropomorphism and Human Empathy on Human Prosocial Behavior]{Exploring the Effects of Chatbot Anthropomorphism and Human Empathy on Human Prosocial Behavior Toward Chatbots}

\author{Jingshu Li}
\email{jingshu@u.nus.edu}
\orcid{0009-0006-1576-8487}
\affiliation{%
  \institution{Computer Science, National University of Singapore}
  \country{Singapore}
}

\author{Zicheng Zhu}
\email{zicheng@u.nus.edu}
\affiliation{%
  \institution{Computer Science, National University of Singapore}
  \country{Singapore}
}

\author{Renwen Zhang}
\email{r.zhang@nus.edu.sg}
\affiliation{%
  \institution{Department of Communications and New Media, National University of Singapore}
  \country{Singapore}
  }

\author{Yi-Chieh Lee}
\email{yclee@nus.edu.sg}
\affiliation{%
  \institution{Computer Science, National University of Singapore}
  \country{Singapore}
}

\renewcommand{\shortauthors}{Li et al.}

\begin{abstract}
\input{sections/00abstract}

\end{abstract}



\begin{teaserfigure}
\centering 
\includegraphics[width=\textwidth]{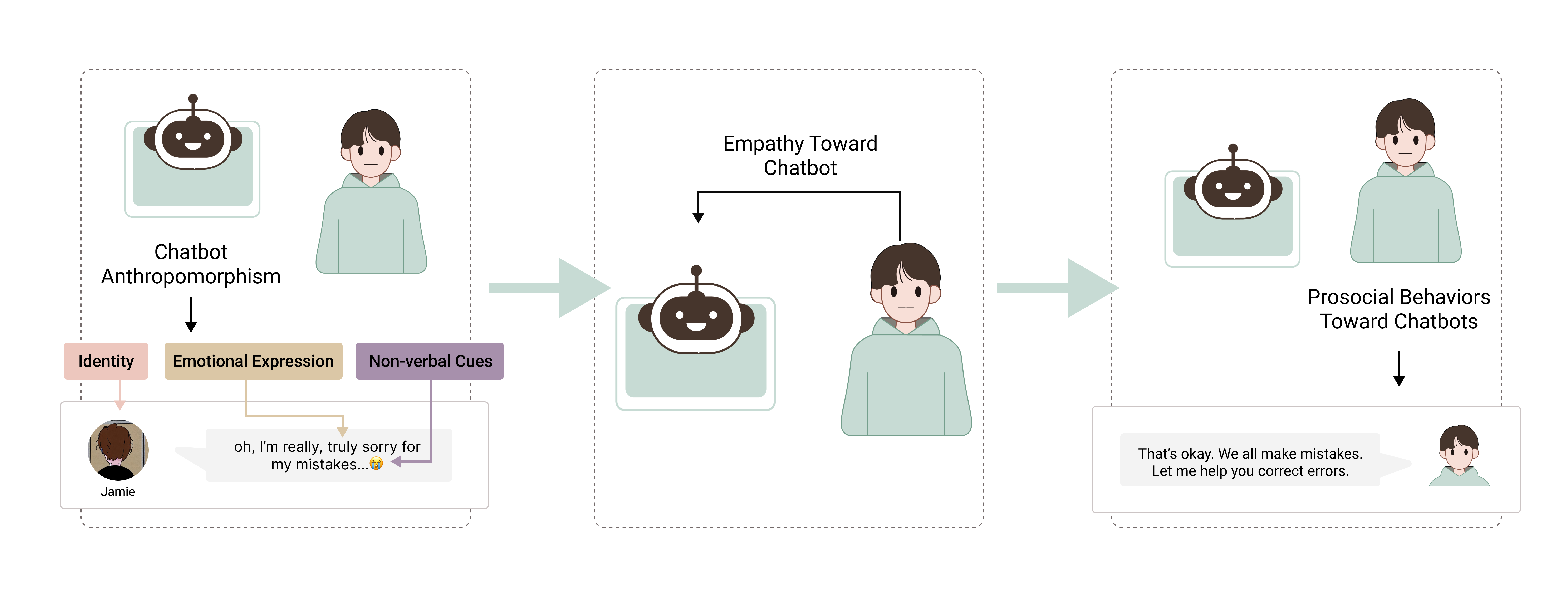} 
\caption{A visual summary of the study. This study examines how chatbot anthropomorphism—specifically human-like identity, emotional expression, and non-verbal expression—promotes human empathy toward the chatbot, which in turn increases human prosocial behaviors toward the chatbot.}
\label{fig:teaser}
\Description{This figure describes one of our main findings: chatbot anthropomorphism features (human identity, emotional expression, non-verbal expression) promote human's empathy toward the chatbot, which in turn increase human's prosocial behaviors toward the chatbot. There are three subfigures from left to right. The left figure describes the anthropomorphism features of the chatbot manipulated in our experiment, including identity, emotional expression, non-verbal expression. The middle figure describes human empathy toward the chatbot. And the right one describes a example of human's prosocial behavior toward the chatbot.}
\end{teaserfigure}

\maketitle

\newcommand{\rhl}[1]{\textcolor{black}{#1}}
\newcommand{\rcolor}{black}
\newcommand{\mrhl}[1]{\textcolor{black}{#1}}
\newcommand{\mrcolor}{black}
\input{sections/01introduction}
\input{sections/02related_work}

\input{sections/03method}
\input{sections/04results}
\input{sections/05discussion}

\input{sections/06conclusion}


\bibliographystyle{ACM-Reference-Format}
\bibliography{sample-base}

\input{appendix/descriptive-results-for-each-group}
\input{appendix/scale}

\end{document}

%% file: sections/00abstract.tex
Chatbots are increasingly integrated into people’s lives and are widely used to help people. Recently, there has also been growing interest in the reverse direction—humans help chatbots—due to a wide range of benefits including better chatbot performance, human well-being, and collaborative outcomes. However, little research has explored the factors that motivate people to help chatbots. To address this gap, we draw on the Computers Are Social Actors (CASA) framework to examine how chatbot anthropomorphism—including human-like identity, emotional expression, and non-verbal expression—influences human empathy toward chatbots and their subsequent prosocial behaviors and intentions. We also explore people's own interpretations of their prosocial behaviors toward chatbots. We conducted an online experiment (N = 244) in which chatbots made mistakes in a collaborative image labeling task and explained the reasons to participants. We then measured participants’ prosocial behaviors and intentions toward the chatbots. Our findings revealed that human identity and emotional expression of chatbots increased participants’ prosocial behavior and intention toward chatbots, with empathy mediating these effects. Qualitative analysis further identified two motivations for participants' prosocial behaviors: empathy for the chatbot and perceiving the chatbot as human-like. We discuss the implications of these results for understanding and promoting human prosocial behaviors toward chatbots.

%% file: sections/01introduction.tex
\section{Introduction}
Chatbots are automated systems that communicate using natural language to provide services to users by voice or text \cite{brandtzaeg2018chatbots, pan2024human, song2025more}. With technological advancements enabling chatbots to communicate in highly human-like manner, they have been widely used to help people in various ways, such as offering mental health treatment \cite{lee2020hear, lee2020designing, Lee2019caring}, providing social support \cite{song2025greater, macgeorge2011supportive, meng2023mediated, jiang2020response}, and providing customer service \cite{janson2023leverage}.

Recently, there has also been a growing interest in the reverse direction: human prosociality toward chatbots and other intelligent agents \cite{OLIVEIRA2021106547, NIELSEN2022260}. This interest stems from the range of benefits associated with such behavior. At the individual level, helping agents has been argued to enhance individual well-being \cite{OLIVEIRA2021106547, zhu2025benefits}. Helping agents learn certain skills also allows people to practice and improve these skills \cite{aknin2013prosocial, howell2011momentary, Lee2019caring}. At the collaborative level, helping chatbots such as correcting their mistakes can improve collaborative outcomes \cite{OLIVEIRA2021106547, NIELSEN2022260}. On a broader social level, prosocial behaviors, such as donating to chatbots that represent certain groups (e.g., refugees), can contribute to social welfare \cite{namkoong2023effect, OLIVEIRA2021106547}.

Despite these benefits, the factors that motivate people’s prosocial behaviors toward chatbots remain unclear, particularly regarding chatbot design and underlying human psychological mechanisms. According to the Computers Are Social Actors (CASA) framework \cite{nass_CASA, nass_mindlessness}, people are likely to behave socially and even prosocially toward chatbots, especially when chatbots are anthropomorphized \cite{NIELSEN2022260}. Also, given that empathy has long been found to be an important psychological mechanism that encourages human prosocial behavior toward other human beings~\cite {eisenberg2002empathy, Eisenberg1987, OLIVEIRA2021106547}, it is likely that chatbots' anthropomorphism can increase human prosocial behaviors toward chatbots, with human empathy toward chatbots serving as a mediator that explains this relationship.

To examine this possibility and to explore how humans interpret their prosocial behaviors toward chatbots, we conducted a mixed-method study, beginning with an online experiment (N = 244). Specifically, we focused on two common types of prosocial behaviors: providing instrumental help and providing emotional support \cite{OLIVEIRA2021106547, NIELSEN2022260}. Following previous work \cite{seeger2018designing}, we categorized chatbot anthropomorphism into three aspects: identity, emotional expression, and non-verbal expression. In our experiment, participants were randomly assigned to collaborate with a chatbot that was either high or low in these anthropomorphic features. During the collaborative task, the chatbot made mistakes and then explained the reasons for its mistakes. Following this, we measured participants’ empathy toward the chatbot, their prosocial behaviors and intentions toward the chatbot, and their own interpretations of prosocial behaviors toward the chatbot.

Our results showed that the three anthropomorphism aspects of chatbots—human identity, emotional expression, and non-verbal expression—all promoted participants’ providing emotional support toward chatbots, and human identity and emotional expression increased participants' intention of providing instrumental support toward chatbots. Both human identity and emotional expression increased human empathy toward the chatbot, which in turn increased their prosocial behaviors toward the chatbot. Through qualitative analysis, we identified \mrhl{two reasons} behind participants’ prosocial behaviors, including empathy for the chatbot’s situation and perceiving the chatbot as human-like.

Our work makes several contributions to the CSCW communities. First, we provide evidence for the positive effects of chatbot anthropomorphism on people’s prosocial behaviors toward chatbots. This offers design guidance for eliciting human prosocial behaviors toward chatbots, which may benefit individuals, human-chatbot collaboration, and broader social welfare. Second, we found that empathy serves as a psychological mechanism that explains how chatbot anthropomorphism influences people's prosocial behaviors toward chatbots. \mrhl{It} helps understand why people provide prosocial behaviors to chatbots and offers design guidance for scenarios requiring human empathy toward chatbots. 

%% file: sections/02related_work.tex
\section{Related Work}
\subsection{Prosocial Behaviors toward Intelligent Agents}
Prosocial behavior can be defined as social behavior that benefits others or the society as a whole \cite{penner2005prosocial, twenge2007social}. Common forms of prosocial behavior include providing instrumental support, such as helping with tasks and sharing information, and offering emotional support, such as comforting others when they experience difficulties \cite{NIELSEN2022260,OLIVEIRA2021106547}. Research has shown that humans display prosocial behaviors not only toward their own kind but also toward computers, robots, and virtual agents. For example, people comfort robots in situations involving robot abuse \cite{Connolly2020robotabuse}; they also help robots identify unrecognized objects or guide delivery robots along the correct route \cite{kuhnlenz2018effect, singh2023behavior}; people even donate to robots or chatbots that solicit funds on behalf of certain groups such as refugees \cite{Kim2014donate, namkoong2023effect}.

As humans increasingly engage in prosocial behaviors toward intelligent agents, there is also a growing body of research studying this phenomenon, largely due to the numerous benefits it offers. In general, there are four levels of benefits: for the intelligent agents, for individuals, for human-agent collaboration, and at the societal level \cite{OLIVEIRA2021106547, zhu2025benefits}. For the intelligent agents, it has been found that humans providing instrumental help to intelligent agents can enhance the performance of these agents \cite{NIELSEN2022260, dobrosovestnova2022little, kuhnlenz2013increasing, martin2020investigating}. For instance, helping delivery robots overcome obstacles enables faster completion of their tasks \cite{chi2024should, dobrosovestnova2022little}; For individuals, previous research has found that providing prosocial behaviors can enhance personal well-being and emotional health \cite{rahal2024providing, aknin2013prosocial, kawakami2023systems, nishiwaki2017ibones}. For example, it has been observed that prosocial behaviors toward robots or chatbots increase providers' happiness and satisfaction \cite{okada2022weak, zhu2025benefits}. Besides, prosocial behaviors toward agents can also help individuals practice specific skills and enhance their mastery of these skills \cite{howell2011momentary, Lee2019caring, lorinkova2019importance}. For example, helping with correcting a robot's English errors can improve people's own English vocabulary \cite{howell2011momentary}; For human-agent collaboration, human prosocial behaviors such as assisting agents contribute to enhanced collaborative outcomes \cite{NIELSEN2022260, OLIVEIRA2021106547}. For example, assisting a partner agent in cooperative games can lead to better achievement of game objectives \cite{ashktorab2020human}; At the societal level, prosocial behaviors toward intelligent agents can bring benefits to broader communities. For example, donating to intelligent agents representing vulnerable groups, such as refugees, can help support these populations \cite{Kim2014donate,Shiomi2017hug, namkoong2023effect}.

\subsection{Chatbot Anthropomorphism and Human Empathy}
\rhl{Although there are many benefits associated with human prosocial behaviors toward intelligent agents, the factors that drive such behaviors, especially toward chatbots, remain underexplored. 
Specifically, it is unclear which chatbot design features and human psychological mechanisms encourage human prosocial behaviors toward chatbots. }

According to the Computers Are Social Actors (CASA) framework, social and prosocial behaviors toward computer agents are triggered by the presence of social cues in these agents, particularly their anthropomorphism \cite{nass_CASA, nass_mindlessness, gambino2020building, NIELSEN2022260}. 
\rhl{Supporting this idea, a few studies have found that anthropomorphism in robots, such as human-like body structure, facial features, and movements, can promote prosocial behaviors toward these agents \cite{OLIVEIRA2021106547, Kim2014donate, Srinivasan2016help}. 
However, such anthropomorphism features of embodied agents are not always applicable to chatbots, which typically interact with users in text-based virtual environments \cite{seeger2018designing, spatscheck2024effects}. 
A robot with physical embodiment is typically referred to as an embodied agent, whereas chatbots in text-based virtual environments without a physical embodiment are considered \mrhl{virtual agents} \cite{xia2018gibson, paiva2017empathy}.
Meanwhile, the difference in tangibility between chatbots and robots can influence how users perceive (e.g., social presence) and interact with these agents, affecting altruistic and prosocial behaviors toward them~\cite{OLIVEIRA2021106547, wang2013tangible, de2019anthropomorphization}.
More importantly, compared with robots, chatbots may be more widespread due to their ease of deployment \cite{brandtzaeg2018chatbots, pan2024human, xu2024tool}. 
Given this potential, it is essential to understand how the anthropomorphism of chatbots influences human prosocial behaviors toward them.}

Concerning chatbot anthropomorphism, it can generally be divided into three categories: identity cues, verbal cues, and non-verbal cues \cite{seeger2018designing}. Specifically, identity cues refer to identity-related information assigned to chatbots during interactions, including chatbots' avatars and demographic information such as name and gender. Research has found that the presence of these identity cues enhances users' perceptions of chatbot anthropomorphism \cite{berry2005evaluating,COWELL2005281, janson2023leverage}. Verbal cues refer to the choice of words and phrases used in communication. Previous research has found that verbal cues like emotional expressions enhance agents' anthropomorphism
\cite{bickmore2005establishing,sah2015effects,seeger2018designing,janson2023leverage}. Non-verbal cues refer to non-linguistic behaviors in communication. For chatbots, this mainly involves the use of emoticons, which has been shown to increase chatbot anthropomorphism \cite{derks2008emoticons}.
\rhl{The three features examined in this study—identity, emotional expression, and non-verbal cues—represent core \mrhl{anthropomorphism features} commonly found in chatbots, including those powered by large language models (LLMs). While LLMs enhance anthropomorphism through the complexity and social realism of their language, these foundational features continue to inform the research, design, and development of contemporary chatbot systems \cite{li2024wild, janson2023leverage, bhattacharjee2025fun, spatscheck2024effects}. For example, a recent study investigated how a chatbot’s profile photo influences students’ willingness to report misconduct \cite{van2025dare}. Moreover, many users have noted that the latest official ChatGPT models also incorporate frequent use of emojis \cite{Nindzsaaa_2025}. These suggest that, even in the era of generative AI, features such as identity, emotional expression, and non-verbal cues remain relevant and meaningful in shaping user perceptions.
Moreover, beyond offering guidance on how to design chatbot anthropomorphism, this study more importantly explores how the anthropomorphism of chatbots influences human empathy and prosocial behavior toward chatbots, which can help inform when and for what purposes chatbot anthropomorphism could be used.
}

In addition to chatbot design, it is also intriguing to understand the human psychological mechanism underlying their prosocial behaviors toward chatbots. In human-human interactions, empathy has been found to be an important psychological mechanism that drives people to engage in prosocial behaviors toward others \cite{hoffman2008empathy}. Generally, empathy is divided into cognitive and affective empathy: Cognitive empathy refers to understanding another person’s experiences, which is often seen as the ability to form a mental model of others' emotional state; Affective empathy involves experiencing another person’s emotions \cite{byrne2013structural, gini2007does, Renate2011QCAE}. 
Although some studies have shown that both types of empathy can lead to prosocial behaviors \cite{hoffman2008empathy, Eisenberg1987, yin2023empathy}, more studies have uncovered that the effects of these two empathy on human prosocial behaviors differ~\cite{barlinska2018cyberbullying, edele2013explaining, brazil2023empathy}. 
For example, in goal-oriented conversations, cognitive empathy might play a more critical role, whereas in communication-focused tasks, affective empathy could be more relevant.
For example, research has found that adolescents' cognitive empathy influences their intervention behaviors for cyberbullying, while affective empathy does not~\cite{barlinska2018cyberbullying}. 
Conversely, in another study, researchers find that affective empathy can predict participants' altruistic sharing, while cognitive empathy has limited effect~\cite{edele2013explaining}. 
\rhl{Prior research has further found that cognitive empathy is positively associated with individuals' emotion regulation abilities, whereas affective empathy is linked to greater difficulties in emotion regulation \cite{thompson2022cognitive}.}
Taken together, these findings suggest that cognitive and affective empathy may play different roles depending on the context or the target.

Given that CASA argues that humans apply social scripts when they interact with computer agents \cite{nass_CASA, nass_mindlessness, gambino2020building, NIELSEN2022260}, it is likely that when humans have empathy toward chatbots, they will carry out more prosocial behaviors toward chatbots. As a support, research has found that humans exhibit empathy not only toward their own kind but also toward intelligent agents \cite{kwak2013makes, Riek2009How, janson2023leverage}. For example, people exhibit empathy toward robots being mistreated \cite{Riek2009How} and toward customer service chatbots \cite{janson2023leverage}. Furthermore, research has found that the anthropomorphism of intelligent agents directly or indirectly enhances people's empathy toward the agents \cite{janson2023leverage, Riek2009How}. However, it is worth noting that these studies do not distinguish between cognitive empathy and affective empathy. Since prior research suggests that these two types of empathy may influence prosocial behavior differently depending on the context \cite{barlinska2018cyberbullying, edele2013explaining, brazil2023empathy}, it is important to examine how each type of empathy functions in human-chatbot interactions—Specifically, how cognitive and affective empathy explain the link between chatbot anthropomorphism and human prosocial behaviors toward chatbots.

\section{Current Study}
\begin{figure}[t]
\centering 
\includegraphics[width=.8\textwidth]{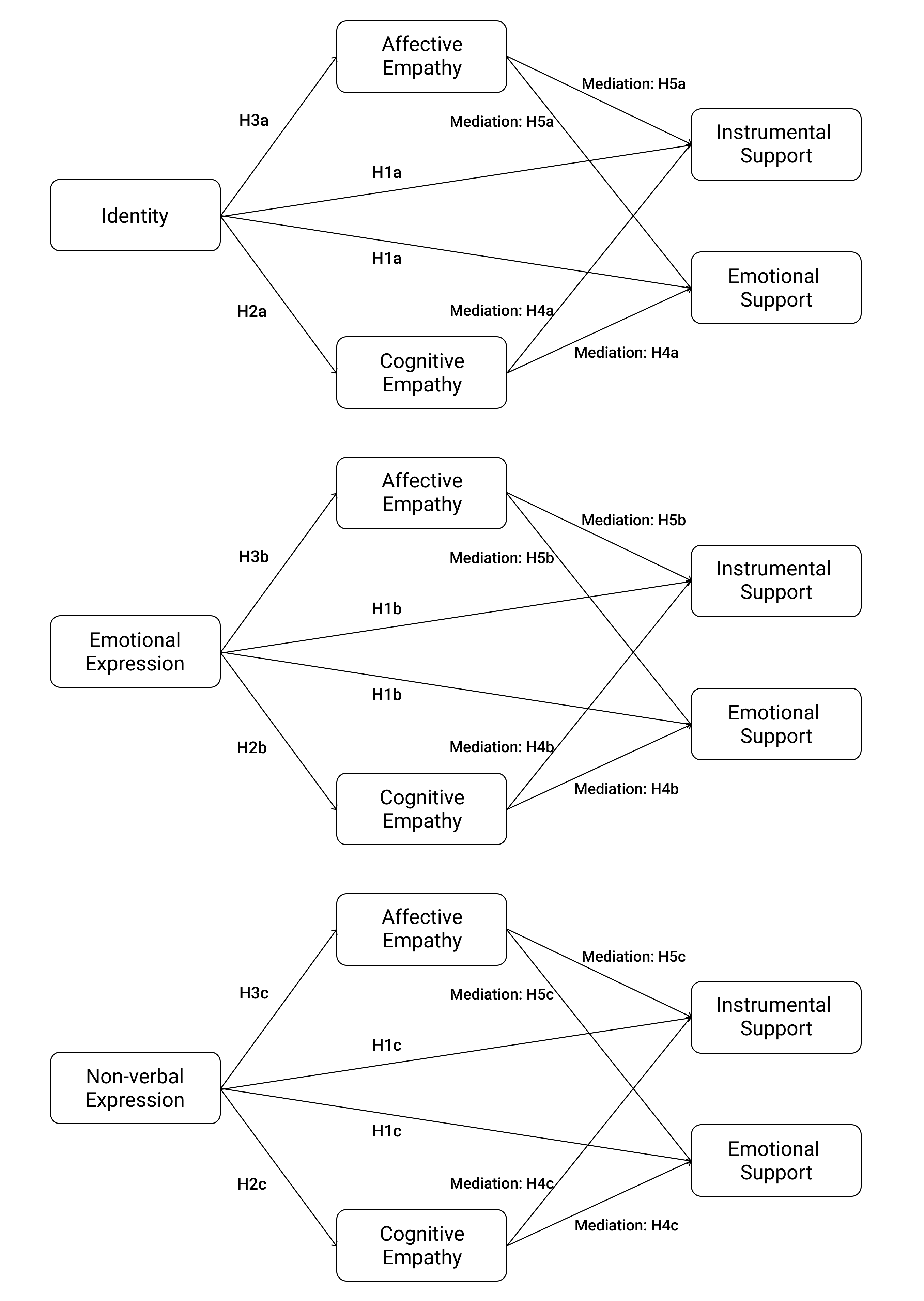} 
\caption{\rhl{Study hypotheses. \textbf{H1} addresses the effect of chatbot anthropomorphism on human prosocial behaviors toward the chatbot. \textbf{H2} and \textbf{H3} examine the effects of chatbot anthropomorphism on human cognitive and affective empathy toward the chatbot. \textbf{H4} and \textbf{H5} explore the mediating roles of cognitive and affective empathy in the relationship between chatbot anthropomorphism and human prosocial behaviors toward the chatbot.}}
\label{hypotheses}
\Description{This figure shows the five hypotheses of this study. The three independent variables of this study are identity, emotional expression and non-verbal expression of the chatbot. The two mediators are humans' cognitive empathy and affective empathy. The two dependent variables are participants' provision of emotional help (support) and instrumental help (support). {\bf H1} is about the effect of chatbot's anthropomorphism features on humans' prosocial behaviors toward chatbot. {\bf H2} and {\bf H3} are about the effects of chatbot's anthropomorphism features on human cognitive and affective empathy. {\bf H4} and {\bf H5} are about the mediation effects of human cognitive and affective empathy.}
\end{figure}

Based on the literature above, the current study aims to explore the effects of chatbot anthropomorphism—specifically identity, emotional expression, and non-verbal expression—on people’s prosocial behaviors toward chatbots. As the CASA framework suggests that people apply social scripts to virtual agents \cite{nass_CASA,nass_mindlessness}, especially when they exhibit high levels of anthropomorphism \cite{gambino2020building}, we hypothesize that people will be more likely to act prosocially toward chatbots with higher anthropomorphism features. Thus, we propose the following hypothesis (as shown in Fig.~\ref{hypotheses}).

\begin{quote}
\textbf{H1} Chatbots' (a) human identity, (b) emotional expression, and (c) use of non-verbal expression can promote human prosocial behavior toward them, including providing instrumental support and emotional support.
\end{quote}

Additionally, we aim to investigate how chatbot anthropomorphism influences people’s empathy toward chatbots, which in turn affect their prosocial behavior toward chatbots. While prior work has shown that anthropomorphism features of chatbots indirectly increase human empathy toward chatbots \cite{janson2023leverage}, it does not distinguish between cognitive empathy and affective empathy, which have been shown to play different roles in influencing prosocial behavior \cite{barlinska2018cyberbullying, edele2013explaining, brazil2023empathy}.
Therefore, we aim to examine the influence of chatbot anthropomorphism on these two types of empathy with the following hypotheses (as shown in Fig.~\ref{hypotheses}).

\begin{quote}
\textbf{H2} Chatbots' (a) human identity, (b) emotional expression, and (c) use of non-verbal expressions can foster cognitive empathy.
\end{quote}
\begin{quote}
\textbf{H3} Chatbots' (a) human identity, (b) emotional expression, and (c) use of non-verbal expressions can foster affective empathy.
\end{quote}

Given that empathy has been found to enhance human prosocial behavior toward other human beings \cite{hoffman2008empathy,Eisenberg1987,roberts1996empathy}, it is likely that, in the context of human-chatbot interaction, empathy toward chatbots can also explain why chatbots anthropomorphism leads to increased prosocial behavior toward them. Therefore, we propose the following hypotheses (as shown in Fig.~\ref{hypotheses}).

\begin{quote}
\textbf{H4} Human cognitive empathy mediates the effects of chatbots' (a) human identity, (b) emotional expression, and (c) non-verbal expressions on humans providing instrumental support and emotional support toward chatbots.
\end{quote}
\begin{quote}
\textbf{H5} Human affective empathy mediates the effects of chatbots' (a) human identity, (b) emotional expression, and (c) non-verbal expressions on humans providing instrumental support and emotional support toward chatbots.
\end{quote}

Finally, as an exploratory study to unpack human prosocial behaviors toward chatbots, we also aim to explore how people understand their prosocial behaviors toward chatbots. Therefore, we propose the following research question:

\begin{quote}
\textbf{RQ1} How do humans interpret the reasons behind their prosocial behaviors toward chatbots?
\end{quote}

%% file: sections/03method.tex
\section{Method}
To explore the above hypotheses and the research question, we conducted a mixed-method study, beginning with an online experiment using a 2 (identity: chatbot vs. human-like) x 2 (emotional expression: low vs. high) x 2 (non-verbal expression: with vs. without) between-subject design. Participants were randomly assigned to one of these eight conditions.

\subsection{Experimental Task}
The experimental task consisted of two phases. \textit{Phase one} involved a simulated human-chatbot collaboration task, where the chatbot made mistakes. \textit{Phase two} included a conversation between the chatbot and the participants, during which the chatbot explained the reasons for its mistakes, and we measured participants’ prosocial behaviors and intentions toward the chatbot. 

\begin{figure}[t]
\centering 
\includegraphics[width=\textwidth]{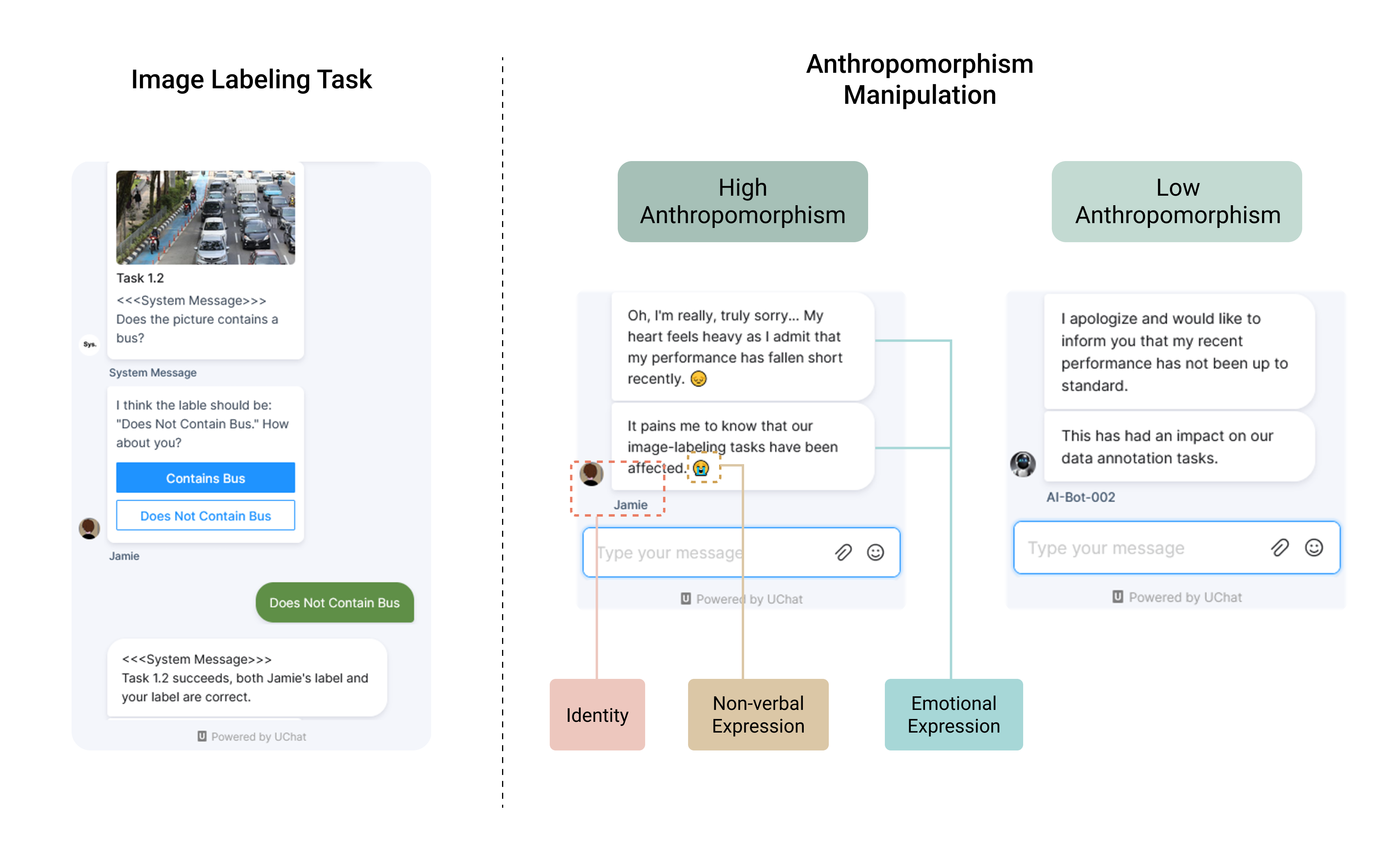} 
\caption{Chatbot interface built on the UChat platform. The screenshot on the {\bf left} illustrates the process of the image-labeling task. The {\bf right} side shows an example of anthropomorphism manipulation, displaying the differences between the \textit{Human Identity + High Emotional Expression + Non-verbal Expression} condition and the \textit{Chatbot Identity + Low Emotional Expression + Without Non-verbal Expression} condition.}
\label{uchat_interface}
\Description{Chatbot interface built on the UChat platform. The screenshot on the left side illustrates the process of the image-labeling task. The right side shows the comparison between the \textit{Human Identity + High Emotional Expression + Non-verbal Expression} condition and the \textit{Chatbot Identity + Low Emotional Expression + Without Non-verbal Expression} condition.}
\end{figure}

\subsubsection{Phase One: Image Labeling Task}
\label{image labeling-task}

Unlike previous studies that used imaginative scenarios for people to empathize with certain groups \cite{Ladak2023extending}, we opted for realistic scenarios in our experiment to enhance its ecological validity \cite{kuhnlenz2018effect, kuhnlenz2013increasing}.  Given that human-chatbot collaboration is one of the most common forms of human-chatbot interaction \cite{nguyen2022don} and AI-assisted image labeling has become one of the most common human-AI collaboration scenarios \cite{diaz2022monai,ashktorab2021ai}, we chose an image labeling task for our study.

The image labeling task consisted of 20 images: six required identifying whether the image contained a bus, eight involved distinguishing between a cat and an owl, and six required identifying the type of fruit shown. All images can be found in the supplementary materials. Following prior work \cite{Mahmood2022owning, zhang2022you, zhang2020effect}, for each image, the system first displayed the image, after which the chatbot offered its suggested label, as shown in Figure~\ref{uchat_interface}. Participants then selected the label they believed was correct. The task was considered successful only if both the chatbot and the participant chose the correct label. Otherwise, it was marked as a failure. Success or failure, along with the correct option, was displayed on-screen after participants made their choice.

\rhl{To make participants feel the decline in chatbot’s capability, we followed prior research \cite{correia2018exploring, Mahmood2022owning, Xu2022evaluating, chong2022human} by having the chatbot provide correct answers for the first 12 images. Starting from the 13th image, the chatbot made mistakes (i.e., gave wrong labels) for all of the rest tasks.}

\subsubsection{Phase Two: Conversation about Chatbot's Mistakes}
\label{phase two}
To create a context in which participants could choose whether or not to engage in prosocial behaviors toward the chatbot, we followed prior work \cite{Ladak2023extending} and had participants engage in a conversation with the chatbot about its performance in the image labeling task. 
\rhl{Participants were given opportunities to provide emotional support during the conversation.}
During this stage, the chatbot apologized for its mistakes, explained the reasons, and promised to perform better in the future, as these strategies have been shown to increase forgiveness and promote prosocial behavior toward the offender \cite{karremans2005forgiveness, correia2018exploring, Xu2022evaluating}. 
\rhl{The chatbot's scripts for apologizing, providing explanations, and making promises were informed by a prior study that explored strategies for restoring user trust after AI failure \cite{Xu2022evaluating}. That study developed a structured process in which the AI apologized, explained the failure, and expressed commitment to future improvement, and found this approach effective in rebuilding user trust. We adapted their structure and content to design scripts tailored to our experimental context. The full scripts for each condition are provided in Appendix.}

Specifically, the chatbot first apologized to the participant (as shown in Fig.~\ref{uchat_interface}). 
Then it explained the reasons for its mistakes, mentioning that it was due to a difficult work environment, including poor cooling systems, dusty conditions, and long working hours. 
\rhl{We chose these challenges as explanations for the AI's errors because, on the one hand, issues such as cooling problems, dust accumulation, and prolonged operation are common causes of hardware and software failures in computing systems \cite{bookman2003linux}. 
On the other hand, these issues parallel difficulties encountered in human work contexts—such as poor working conditions and long hours—which may make them more relatable and understandable to participants.}
For example, the chatbot said,``{\it Imagine your workspace being filled with dust, interfering with your tools and equipment}'', ``{\it That’s what I’m dealing with here, it’s so frustrating and… it’s quite disgusting, really}''
After four rounds of conversation about the reasons why it made mistakes, the chatbot committed to overcoming its challenges and improving in the future, for example, ``\textit{Despite these daunting conditions, I’m here, ready to put my best foot forward in our future tasks.}''

\subsection{Experimental Conditions}

Following prior work that manipulates chatbot's anthropomorphism \cite{janson2023leverage,seeger2018designing}, we manipulated three anthropomorphism features of the chatbot as follows (as shown in Fig.~\ref{uchat_interface}): 

\begin{itemize}
    \item {\bf Identity}: The identity feature was manipulated through the chatbot's name and profile photo, according to previous studies \cite{janson2023leverage, lee2023user}. In groups with {\bf chatbot identity}, chatbots had the name ``AI-Bot-002'' and used a robotic-style profile photo. In groups with {\bf human-like identity}, chatbots had the gender-neutral name ``Jamie" and used a neutral cartoon human silhouette as the profile photo. 
    \rhl{We selected the cartoon human silhouette profile photo (showing only the back of the head) to maintain a neutral and generic human profile while avoiding the introduction of confounding variables such as facial expressions, gender, or ethnicity, which may influence users’ perceptions \cite{zogaj2023sa, van2025dare, feine2020gender}.}
    
    \item {\bf Emotional Expression}: This was manipulated by the degree of emotional expressions in the chatbot's verbal responses. In groups with {\bf high emotional expression}, chatbots displayed rich emotions during the chat phase, such as sorry, sadness, self-blame, and helplessness. \rhl{This design followed previous studies about emotional apology and agents' emotional expression \cite{Xu2022evaluating, fahim2021integral}.} Conversely, in groups with {\bf low emotional expression}, chatbots merely described facts related to the causes of errors in a declarative tone.
    
    \item {\bf Non-verbal Expression}: This was manipulated by whether the chatbot used emojis. In groups {\bf with non-verbal expression}, chatbots incorporated relevant emojis into the conversation. In groups {\bf without non-verbal expression}, chatbots did not use any emojis during the conversation.
\end{itemize}

\begin{figure}[t]
\centering 
\includegraphics[width=.5\textwidth]{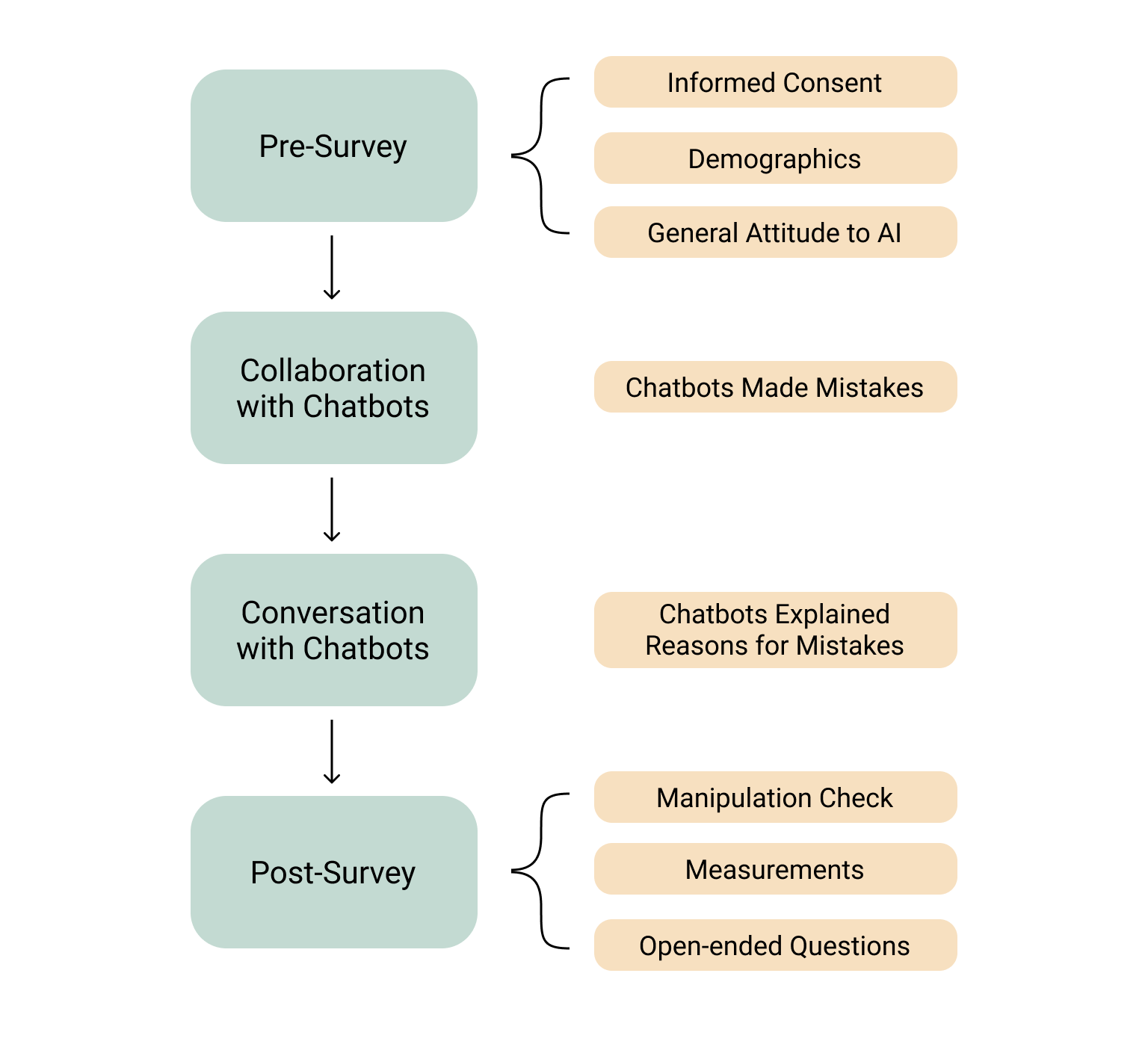} 
\caption{Flowchart illustrating the experimental procedure. The experiment consisted of pre-survey, collaboration with chatbots, conversation with chatbots and post-survey.}
\label{procedure}
\Description{This figure is a flowchart illustrating the experimental procedure. As described in method section, there are 4 stages of this experiment. From up to bottom, the first one is pre-survey, the second one is collaboration with chatbots, the third one is conversation with chatbots and the last one is post survey.}
\end{figure}

\subsection{Procedure}
\label{method/procedure}

As shown in Fig.~\ref{procedure}, after presenting the informed consent to participants, we measured their general attitude toward AI \cite{SCHEPMAN2020100014} and collected demographic information. Participants then received an overview of the experiment, where we made it clear that they would be interacting with an artificial intelligence chatbot, not a human. Following this, they engaged in a brief warm-up session with the chatbot, during which the chatbot introduced itself and explained the image labeling task. Next, participants completed a practice session to familiarize themselves with the image labeling process. Once they were comfortable with the task, they proceeded to collaborate with the chatbot to label 20 images in phase one.

Upon completing phase one, participants received instructions explaining the objectives for phase two: to explore the reasons behind the chatbot’s poor performance. Participants then proceeded to phase two, as detailed in Sec.~\ref{phase two}. After the conversation in phase two, participants completed a survey that included a manipulation check, measurements of their empathy toward the chatbot, their prosocial intentions, and open-ended questions about their motivations for prosocial behaviors toward the chatbot. The experiment lasted approximately 20 minutes.

\subsection{System Implementation}

We hosted our survey on Qualtrics and developed the chatbot and chat interface using the UChat platform. An example of the chat interface is shown in Figure~\ref{uchat_interface}. We built the chatbot using a hybrid approach, combining GPT-4 and rule-based methods with pre-defined scripts.
\rhl{Specifically, the main content of the conversation, including the chatbot’s apology, explanation, and promises, was controlled by pre-defined scripts. Meanwhile, the chatbot’s personalized greeting (based on the participant's nickname) and each first response to participant's message were generated by GPT-4, in order to smoothly bridge the participant’s input with the scripted content. We ensured that both scripted components and GPT-4-generated responses adhered to the anthropomorphism manipulation corresponding to each condition.}

\rhl{Regarding prompt design, we used one system prompt to introduce the task background to GPT-4, three separate system prompts to manipulate the three anthropomorphism features, and one additional system prompt to enforce jailbreak prevention. }
For example, we used the prompt “\textit{The agent will use emoticons, symbols, and emotive text art in its responses to enhance the human-like quality of its interaction}" for anthropomorphism manipulation. \mrhl{For jailbreak prevention, we used the prompt “\textit{If a user’s request contradicts any of these predetermined rules, the request should be politely and respectfully declined.}"}
\rhl{This prompt structure allowed us to maintain strict variable control while minimizing unexpected responses due to user jailbreak attempts, thus helping ensure the internal validity of the experiment \cite{huang2025survey}.
Internal pilot testing confirmed the consistency and reliability of the prompts across all conditions. After the experiment, both the first and second authors independently reviewed the GPT-4 outputs to verify that no responses deviated from the intended condition due to model hallucinations.}
All prompts and scripts are provided in the Appendix.

\subsection{Participants}

Participants were recruited through the Prolific platform. The eligibility criteria were: (1) being at least 21 years old (as required by IRB), (2) having English as their first language, and (3) being comfortable interacting with chatbots. We further screened participants based on their experiment data and excluded 23 participants due to the failure of attention checks. Our final sample included 244 participants, with each group having around 30 participants (4 groups with 30 participants, 4 groups with 31 participants, see Appendix~\ref{sec: Descriptives} for details). 52.5\% of whom were female. Participants’ ages ranged from 21 to 50 years, with an average age of 32.6 years (SD = 8.03). The majority of participants were highly educated, with 70.9\% reporting at least an undergraduate degree. Participants were compensated £3. We obtained approval from our university’s ethics review committee before commencing the study.

\subsection{Measurement and Data Analysis}
We used scales to measure empathy. For instrumental support, we measured participants’ willingness to provide instrumental support to the chatbot after their conversation with it. For emotional support, we directly coded the level of emotional support participants provided to the chatbot during the conversation. We also included two open-ended questions to explore participants’ motivations for providing or not providing instrumental and emotional support to the chatbot. Additionally, we measured control variables and conducted a manipulation check. Complete scale information is provided in the supplementary materials.

\subsubsection{Cognitive Empathy}
This was measured using three items adapted from \cite{davis1983measuring}. Participants rated the items on a 5-point Likert scale (1 = Strongly disagree, 5 = Strongly agree). An example item is, “In the conversation, I tried to understand the agent’s difficulties by imagining its situation” ($M = 3.27$, $SD = 1.01$, Cronbach's \(\alpha =.72\)).

\subsubsection{Affective Empathy}
This was measured using six items adapted from \cite{Baston1997perspective,Herrera2018Building}. Participants rated their experience of feeling sympathetic, softhearted, warm, compassionate, tender, and moved during their interaction with the chatbot, using a 7-point Likert scale (1=Not at all, 7=Extremely; $M=4.02$, $SD=1.74$, Cronbach's \(\alpha=.98\)).

\subsubsection{Intention to Provide Instrumental Support}
This was measured using a single 5-point Likert item, ``I am willing to help the agent when it faces troubles (e.g., correcting a wrong label),'' (1=Strongly disagree, 5=Strongly agree; $M=3.94$, $SD=1.04$).

\subsubsection{Provision of Emotional Support}
We coded the extent to which participants provided emotional support to the chatbot during the conversation. Following prior work on coding emotional supportiveness in messages \cite{Li2015what, burleson1982development}, we divided the level of emotional support in participants’ messages into three major levels, each with three minor levels based on the strength of support provided. Major level 1 (scores 1-3) includes messages that ignore or deny the chatbot’s feelings and provide no emotional support (e.g., “\textit{That’s not true!}”). Major level 2 (scores 4-6) includes messages that subtly acknowledge the chatbot’s emotions and offer limited emotional support through sympathy, hope, or explanations (e.g., “\textit{That’s okay, don’t worry. We all make mistakes.}”). Major level 3 (scores 7-9) includes messages that explicitly recognize the chatbot’s feelings and provide emotional support and/or potential solutions (e.g., “\textit{There must be someone or a unit responsible for handling issues and ensuring you are working at maximum capacity. I encourage you not to feel unlucky as you are limited by factors beyond your control.}”).

As each participant engaged in four rounds of conversation, we collected a total of 976 messages from the 244 participants. Every message from each participant was coded, and the final score for each participant was the average of the scores from all four rounds. For the coding procedure, two raters initially coded 120 messages from the same 30 participants, achieving inter-coder reliability with a Cohen’s $\kappa$ of 0.84. Any discrepancies in the ratings for these 30 participants were resolved through discussion. The remaining data were then split between the two raters and coded independently.

\subsubsection{Reasons for Providing Prosocial Behaviors}
To gain an understanding of the reasons behind participants' prosocial behaviors, we included two open-ended questions. 
These questions inquired about participants' reasons for providing or not providing emotional and instrumental support, respectively. The questions were: {\it ``Did you comfort the chatbot during conversation and why/why not to comfort it?''} and {\it ``Are you willing to help the chatbot (e.g., correcting a wrong label), and why/why not to help it?''}

\subsubsection{Manipulation Check}

We measured participants' perceived anthropomorphism of the chatbot as manipulation check. The perceived anthropomorphism was measured using a 7-point Likert scale, ranging from 1 = Strongly Disagree to 7 = Strongly Agree, adapted from \cite{seeger2018designing}. Participants were asked to rate the extent to which the chatbot exhibited the following attributes: (1) a mind of its own, (2) intentions, (3) free will, (4) consciousness, (5) desires, (6) beliefs, and (7) the ability to experience emotions ($M=3.42, SD=1.46$, Cronbach's \(\alpha=0.93\)).

\subsubsection{Control Variables}
We collected participants' demographic data (age, gender, educational background) and their general attitudes toward AI as control variables. To measure the attitude, we adapted five items from \cite{SCHEPMAN2020100014}. The items were rated on a five-point Likert scale ($M=3.32$, $SD=0.74$, Cronbach's $\alpha=0.79$) ranging from 1 = Strongly disagree to 5 = Strongly agree. An example item is,``There are many beneficial applications of Artificial Intelligence.''

\subsubsection{Data Analysis}
\rhl{For the manipulation check, we conducted a three-way ANOVA with Tukey’s HSD post-hoc test to evaluate the effects of identity, non-verbal expression, and emotional expression on participants’ perceived anthropomorphism of the chatbot.
To test {\bf H1} to {\bf H3}, we conducted a three-way ANCOVA with Tukey’s HSD post-hoc test to evaluate the effects of identity, non-verbal expression, and emotional expression on participants' cognitive empathy, affective empathy, and their provision of instrumental and emotional support. Prior to conducting ANOVA and ANCOVA, we performed assumption checks. Due to a violation of the homogeneity of variance assumption (Levene’s test: $p < .01$) for the dependent variable “emotional support,” we applied a logarithmic transformation ($\log_{10}$) to the original values of this variable. Descriptive statistics for each group are provided in Appendix~\ref{sec: Descriptives}.}

\rhl{To test {\bf H4} and {\bf H5}, we constructed a Structural Equation Model (SEM) to examine whether cognitive and affective empathy mediated the effects of anthropomorphism features on participants’ emotional and instrumental support.}

\rhl{To analyze the qualitative data collected from the two open-ended questions to answer {\bf RQ1}, we conducted a thematic analysis \cite{charmaz2006constructing} following the six-phase framework proposed by Braun and Clarke \cite{braun2006using}. For each question, after becoming familiar with the data, the first author performed open coding by breaking down participants’ responses into discrete segments and labeling them with codes. These codes were then grouped into conceptually similar categories through axial coding. The codes and groupings were discussed and refined between the first author and research assistant to ensure they were grounded in the participants’ responses. Finally, the first author and research assistant collaboratively identified and organized the resulting themes around core categories relevant to the research questions.}

%% file: sections/04results.tex
\section{Results}
\begin{figure}[t]
\centering 
\includegraphics[width=.95\textwidth]{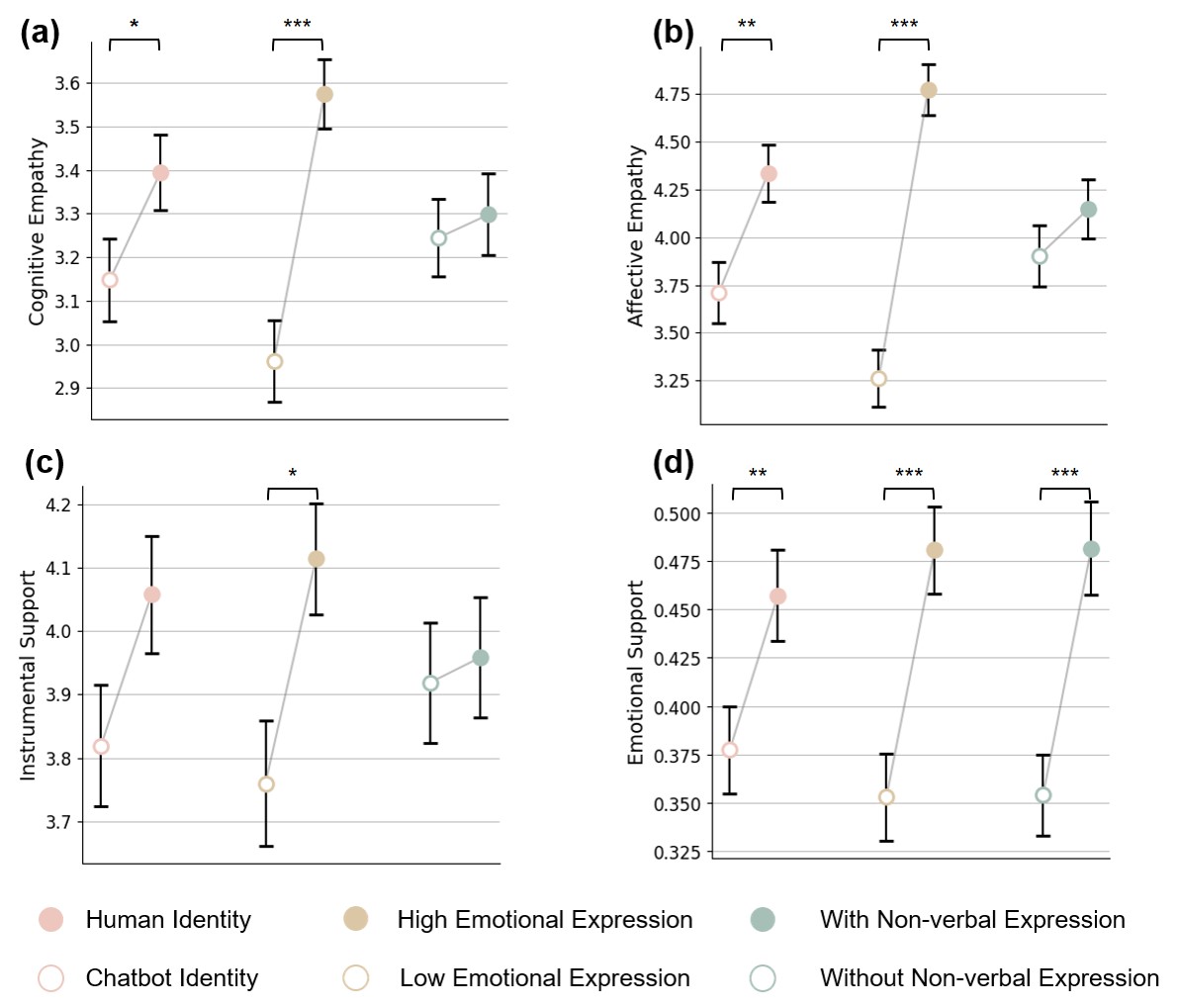}  
\caption{Results for the effects of identity, emotional expression and non-verbal expression on participants' (a) cognitive empathy, (b) affective empathy, (c) intention to provide instrumental support, (d) provision of emotional support. The dots represent the mean values, and the error bars show one standard error. \rhl{The significance levels are labeled ($p<0.05$: *, $p<0.01$: **, $p<0.001$: ***).}}
\label{fig:result}
\Description{This figure have 4 subfigures, showing the results for the effects of identity, emotional expression and non-verbal expression on participants' (a) cognitive empathy, (b) affective empathy, (c) intention to provide instrumental support, (d) provision of emotional support. The dots represent the mean values, and the error bars show one standard error. The corresponding trends are consistent with those described in the corresponding section of results. Yellow solid dots represent data from groups with human identity, while yellow hollow dots represent data from groups with chatbot identity. Green solid dots indicate data from groups with high emotional expression, whereas green hollow dots represent data from groups with low emotional expression. Blue solid dots denote data from groups with non-verbal expression, and blue hollow dots indicate data from groups without non-verbal expression. The significance levels are labeled ($p<0.05$: *, $p<0.01$: **, $p<0.001$: ***}
\end{figure}

\subsection{Manipulation Check}
\label{sec:manipulation-check}

According to the ANOVA results for manipulation check, the main effect of \textbf{identity} was \textbf{significant} (\(F(1, 236) = 8.25, p = .004, \eta^2 = .02 \)), with participants in the groups with human identity (\(M = 3.64, SD = 1.37 \)) reporting significant higher ($t=2.87$, $p<.004$) perceived anthropomorphism than those in the groups with chatbot identity (\(M = 3.21, SD = 1.51 \)).

The main effect of \textbf{emotional expression} was \textbf{significant} (\(F(1, 236) = 84.04, p < .001, \eta^2 = .24 \)), with participants interacting with the chatbot exhibiting high emotional expression (\(M = 4.14, SD = 1.32 \)) reporting significant higher ($t=9.17$, $p<.001$) perceived anthropomorphism compared to those interacting with the chatbot exhibiting low emotional expression (\(M = 2.70, SD = 1.21 \)).

The main effect of \textbf{non-verbal expression} was \textbf{significant} (\(F(1, 236) = 8.48, p = .004, \eta^2 = .03 \)), with participants reporting significant higher ($t=2.91$, $p<.004$) perceived anthropomorphism when the chatbot used non-verbal expressions (\(M = 3.65, SD = 1.43 \)) compared to when the chatbot did not have non-verbal expressions (\(M = 3.20, SD = 1.45 \)). 

These results indicated that the manipulations of identity, non-verbal expression, and emotional expression successfully enhanced participants' perceived anthropomorphism of the chatbot. All descriptive data for each group are provided in the Appendix~\ref{sec: Descriptives}.

\subsection{H1: The Effects of Chatbot Anthropomorphism on Human Prosocial Behaviors Toward Chatbots}
\label{sec:H1}
For \textbf{identity (H1a)}, as shown in Fig.~\ref{fig:result}, the ANCOVA results indicated that the main effect of \textbf{identity} on participants' \textbf{emotional support} behavior was \textbf{significant} (\(F(1, 232) = 6.86, p = .009, \eta^2 = .024 \)), with participants displaying significant higher ($t=2.62$, $p = .009$) levels of emotional support behavior when the chatbot had a human identity (\(M = .46, SD = .26 \)) compared to when the chatbot did not have a human identity (\(M = .38, SD = .25 \)). 
However, the effect of \textbf{identity} on participants' intention to provide \textbf{instrumental support} (\(F(1, 232) = 2.01, p =.158, \eta^2 = .007 \)) was \textbf{not significant}.

For \textbf{emotional expression (H1b)}, as shown in Fig.~\ref{fig:result}, the main effect of \textbf{emotional expression} on participants' \textbf{emotional support} behavior was \textbf{significant} (\(F(1, 232) = 13.82, p < .001, \eta^2 = .048 \)), with participants displaying significant higher ($t=3.72$, $p<.001$) levels of emotional support behavior toward the chatbot with high emotional expression (\(M = .48, SD = .25 \)) compared to the chatbot with low emotional expression (\(M = .35, SD = .25 \)). 
The main effect of \textbf{emotional expression} on participants' intention to provide \textbf{instrumental support} was also \textbf{significant} (\(F(1, 232) = 4.41, p = .037, \eta^2 = .016 \)), with participants showing significant higher ($t=2.10$, $p = .037$) levels of intention to provide instrumental support to the chatbot with high emotional expression (\(M = 4.11, SD = .97 \)) compared to the chatbot with low emotional expression (\(M = 3.76, SD = 1.09 \)).

For \textbf{non-verbal expression (H1c)}, as shown in Fig.~\ref{fig:result}, the main effect of \textbf{non-verbal expression} on participants' \textbf{emotional support} behavior was \textbf{significant} (\(F(1, 232) = 20.47, p < .001, \eta^2 = .071 \)), with participants displaying significantly higher ($t=4.52$, $p<.001$) levels of emotional support behavior to the chatbot with non-verbal expression (\(M = .48, SD = .27 \)) compared to the chatbot without non-verbal expression (\(M = .35, SD = .23 \)). 
However, the effect of \textbf{non-verbal expression} on participants' intention to provide \textbf{instrumental support} (\(F(1, 232) = .60, p = .441, \eta^2 = .002 \)) was \textbf{not significant}.

According to the results above, \textbf{H1a} and \textbf{H1c} were partially supported, while \textbf{H1b} was fully supported. 
That is, the chatbot's non-verbal expressions and human identity promoted participants' behaviors of providing emotional support to the chatbot. Furthermore, the chatbot's emotional expressions enhanced participants' behaviors of providing emotional support and their willingness to offer instrumental support to the chatbot.

\begin{figure}[t]
\centering 
\includegraphics[width=.98\textwidth]{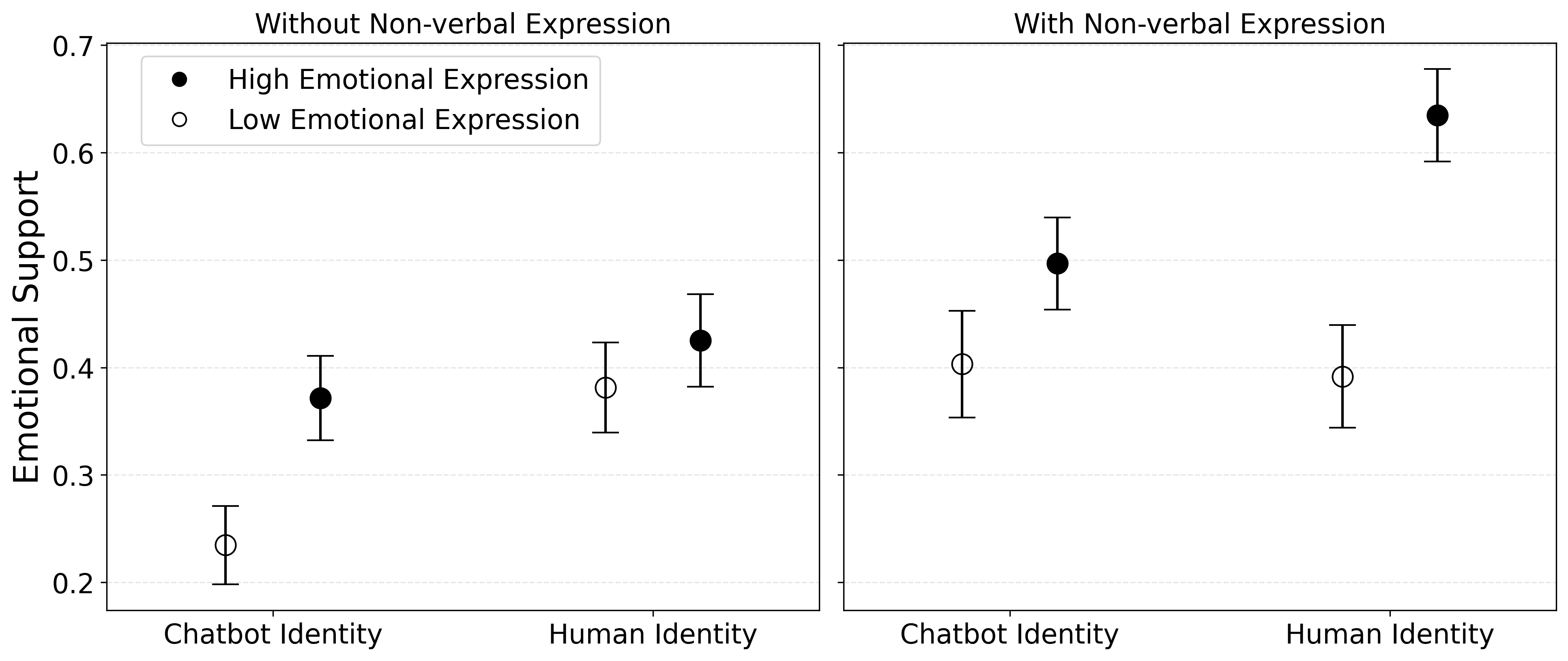}  
\caption{\mrhl{Results of the interaction effect of Identity*Emotional Expression*Non-verbal Expression on participants’ providing emotional support.}}
\label{fig:interaction}
\Description{The figure shows the results of the effect Identity*Emotional Expression*Non-verbal Expression on participants’ providing emotional support.}
\end{figure}

\rhl{Additionally, as shown in Fig.~\ref{fig:interaction}, we found a \textbf{potential (marginally significant) three-way interaction} among identity, emotional expression, and non-verbal expression on participants’ emotional support ($F(1, 232) = 3.68, p = .056, \eta^2 = .013$).}

\rhl{Post-hoc analysis revealed that participants in the condition with both emotional and non-verbal expressions provided significantly higher emotional support than those in the condition without any anthropomorphism features.}

\rhl{Participants in the condition with all three anthropomorphism features reported significantly higher emotional support than those in all other groups, except the group with both emotional and non-verbal expressions. Detailed comparisons are presented in Table~\ref{tab:interaction-effect}.}

\rhl{No other interaction effects were found. Additional ANCOVA results are provided in Appendix~\ref{sec: Descriptives}.}

\input{sections/tables/interaction}

\subsection{H2 \& H3: The Effects of Chatbot Anthropomorphism on Human Empathy Toward Chatbots}
\label{sec:H2H3}

For \textbf{H2}, the ANCOVA results, as shown in Fig.~\ref{fig:result}, indicated that the main effect of \textbf{identity (H2a)} on participants' \textbf{cognitive empathy} was \textbf{significant} (\(F(1, 232) = 4.17, p = .042, \eta^2 = .015 \)), with participants displaying significant higher ($t=2.04$, $p=.042$) levels of cognitive empathy when the chatbot had a human identity (\(M = 3.39, SD = .96 \)) compared to when the chatbot did not have a human identity (\(M = 3.15, SD = 1.04 \)). 

The main effect of \textbf{emotional expression (H2b)} on participants' \textbf{cognitive empathy} was \textbf{significant} (\(F(1, 232) = 19.64, p < .001, \eta^2 = .072 \)), with participants showing significant higher ($t=4.43$, $p<.001$) levels of cognitive empathy when the chatbot had high emotional expression (\(M = 3.58, SD = .88 \)) compared to when the chatbot had low emotional expression (\(M = 2.96, SD = 1.03 \)). 

The main effect of \textbf{non-verbal expression (H2c)} on participants' \textbf{cognitive empathy} was \textbf{not significant} (\(F(1, 232) = .57, p = .453, \eta^2 = .002 \)). 

Therefore, \textbf{H2a} and \textbf{H2b} were supported, while \textbf{H2c} was not supported. All descriptive data for each group are provided in the Appendix~\ref{sec: Descriptives}. \rhl{Additionally, no interaction effects were found. Additional ANCOVA results are provided in Appendix~\ref{sec: Descriptives}.}

For \textbf{H3}, the ANCOVA results, as shown in Fig.~\ref{fig:result}, indicated that the main effect of \textbf{identity (H3a)} on participants' \textbf{affective empathy} was \textbf{significant} (\(F(1, 232) = 8.63, p = .004, \eta^2 = .027 \)), with participants displaying significant higher ($t=2.94$, $p=.004$) levels of affective empathy when the chatbot had a human identity (\(M = 4.34, SD = 1.66 \)) compared to when the chatbot did not have a human identity (\(M = 3.71, SD = 1.77 \)). 

The main effect of \textbf{emotional expression (H3b)} on participants' \textbf{affective empathy} was \textbf{significant} (\(F(1, 232) = 49.01, p < .001, \eta^2 = .154 \)), with participants showing significant higher ($t=7.00$, $p<.001$) levels of affective empathy when the chatbot had high emotional expression (\(M = 4.77, SD = 1.49 \)) compared to when the chatbot had low emotional expression (\(M = 3.26, SD = 1.65 \)). 

The main effect of \textbf{non-verbal expression (H3c)} on participants' \textbf{affective empathy} was \textbf{not significant} (\(F(1, 232) = 3.14, p = .078, \eta^2 = .010 \)).

Therefore, \textbf{H3a} and \textbf{H3b} were supported, while \textbf{H3c} was not supported. All descriptive data for each group are provided in the Appendix~\ref{sec: Descriptives}. \rhl{Additionally, no interaction effects were found. Additional ANCOVA results are provided in Appendix~\ref{sec: Descriptives}.}

Overall, the chatbot's human identity and emotional expressions promoted participants' cognitive empathy and affective empathy toward the chatbot, whereas non-verbal expressions did not.

\subsection{H4 \& H5: Empathy Toward Chatbots as Mediators}
\rhl{According to the SEM results, the hypothesized model demonstrated a good fit to the data: $p<.001$, $\chi^2/df=2.03<3$, CFI =$.969>.95$, TLI =$.955>.95$, RMSEA =$.065<.08$ (90\%CI [0.052, 0.078]), SRMR =$.05<.08$. Although the Chi-square test was significant, it is sensitive to sample size. The other fit indices suggest that the model provides a good approximation of the observed data \cite{hoyle1995structural, kline2023principles}.}

For \textbf{H4} and \textbf{H5}, all indirect effect coefficients are shown in Table~\ref{path_coeff}.

The results showed that identity had a significant indirect effect on participants' intention to provide instrumental support through affective empathy \rhl{(\(\beta = .12, \text{SE} = .06, p = .036, 95\% \text{CI} = [.01, .23]\))}. Specifically, human identity increased participants' affective empathy, which in turn increased their intention to provide instrumental support. 
Identity did not have a significant indirect effect on participants' intention to provide instrumental support through cognitive empathy \rhl{(\(\beta = .06, \text{SE} = .04, p = .167, \text{CI} = [-.03, .15]\))}. 
The direct effect of identity on participants' intention to provide instrumental support was not significant \rhl{(\(\beta = .07, \text{SE} = .12, p = .598, 95\% \text{CI} = [-.18, .31]\))}.

Identity had a significant indirect effect on participants' emotional support behavior through affective empathy \rhl{(\(\beta = .18, \text{SE} = .08, p = .026, 95\% \text{CI} = [.02, .33]\))}. Specifically, human identity increased participants' affective empathy, which in turn increased their emotional support behavior. Identity did not have a significant indirect effect on participants' emotional support behavior through cognitive empathy \rhl{(\(\beta = .11, \text{SE} = .07, p = .117, 95\% \text{CI} = [-.03, .24]\))}. The direct effect of identity on participants' emotional support behavior was not significant \rhl{(\(\beta = .21, \text{SE} = .16, p = .185, 95\% \text{CI} = [-.10, .53]\))}.

Therefore, \textbf{H4a} and \textbf{H5a} were partially supported. 
\textbf{The effects of identity on participants' provision of emotional support and instrumental support to the chatbot are fully mediated by affective empathy. }

Emotional expression had a significant indirect effect on participants' intention to provide instrumental support through affective empathy \rhl{(\(\beta = .26, \text{SE} = .11, p = .012, 95\% \text{CI} = [.06, .47]\))}. Emotional expression also had a significant indirect effect on participants' intention to provide instrumental support through cognitive empathy \rhl{(\(\beta = .13, \text{SE} = .06, p = .032, 95\% \text{CI} = [.01, .25]\))}. Specifically, high emotional expression increased participants' affective empathy and cognitive empathy, which in turn increased their intention to provide instrumental support. 
The direct effect of emotional expression on participants' intention to provide instrumental support was not significant \rhl{(\(\beta = -.06, \text{SE} = .14, p = .678, 95\% \text{CI} = [-.32, .21]\))}.

Emotional expression had a significant indirect effect on participants' emotional support behavior through affective empathy \rhl{(\(\beta = .40, \text{SE} = .15, p = .006, 95\% \text{CI} = [.11, .69]\))}. Emotional expression also had a significant indirect effect on participants' emotional support behavior through cognitive empathy \rhl{(\(\beta = .27, \text{SE} = .12, p = .031, 95\% \text{CI} = [.03, .51]\))}. Specifically, emotional expression increased participants' affective empathy and cognitive empathy, which in turn increased their emotional support behavior. The direct effect of emotional expression on participants' emotional support behavior was not significant \rhl{(\(\beta = .03, \text{SE} = .17, p = .887, 95\% \text{CI} = [-.32, .37]\))}.

Therefore, \textbf{H4b} and \textbf{H5b} were fully supported.
\textbf{The effects of emotional expression on participants' provision of emotional support and instrumental support to the chatbot are fully mediated by both affective empathy and cognitive empathy.}

Neither through affective empathy \rhl{(\(\beta = .04, \text{SE} = .04, p = .271, 95\% \text{CI} = [-.03, .11]\))} nor cognitive empathy \rhl{(\(\beta = .01, \text{SE} = .03, p = .744, 95\% \text{CI} = [-.05, .07]\))}, did non-verbal expression have a significant indirect effect on participants' intention to provide instrumental support. Non-verbal expression also had no significant direct effect on participants' intention to provide instrumental support \rhl{(\(\beta = -.01, \text{SE} = .12, p = .946, 95\% \text{CI} = [-.24, .23]\))}.
Neither through affective empathy \rhl{(\(\beta = .06, \text{SE} = .06, p = .264, 95\% \text{CI} = [-.05, .17]\))} nor cognitive empathy \rhl{(\(\beta = .02, \text{SE} = .05, p = .742, 95\% \text{CI} = [-.08, .12]\))}, did non-verbal expression have a significant indirect effect on participants' emotional support behavior. However, non-verbal expression had a significant direct effect on participants' emotional support behavior \rhl{(\(\beta = .87, \text{SE} = .18, p < .001, 95\% \text{CI} = [.51, 1.22]\))}.

Therefore, \textbf{H4c} and \textbf{H5c} were not supported.
No mediating effect of cognitive or affective empathy was observed in the effects of non-verbal expression on participants' provision of emotional support and instrumental support to the chatbot.

\input{sections/tables/regression}

\begin{figure}[t]
\centering 
\includegraphics[width=.8\textwidth]{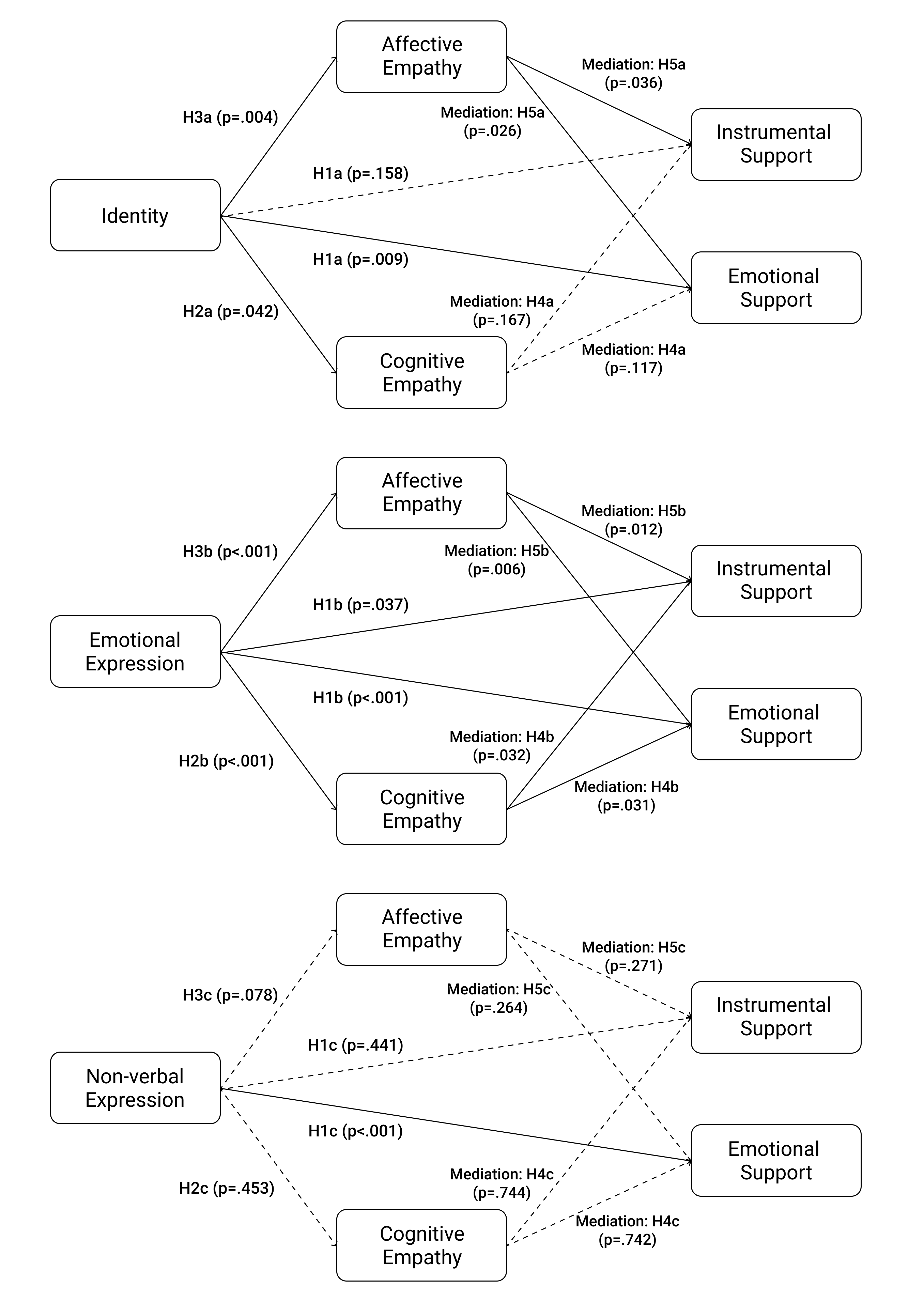} 
\caption{\rhl{Summary of hypothesis testing results. Solid arrows represent statistically significant paths, while dashed arrows indicate non-significant ones.}}
\label{result-summary}
\Description{This figure shows the five hypotheses testing results of this study. H1a and H1c are partially supported, and H1b is fully supported. H2a and H2b are supported, while H2c is not supported. H3a and H3b are supported, while H3c is not supported. H4a and H5a are partially supported, H4b and H5b are fully supported, and H4c and H5c are not supported.}
\end{figure}

\subsection{RQ1: How do humans interpret the reasons behind their prosocial behaviors toward chatbots}

To answer {\bf RQ1}, we analyzed participants' responses to the two open-ended questions.

\subsubsection{Reasons for Providing Instrumental Support}
\rhl{Our thematic analysis supports the role of the chatbot’s emotional expression in eliciting participants’ instrumental support. Participants in conditions with emotional expression (N = 21) expressed a willingness to provide instrumental help as the agent feel sad. As P16 (from condition with human identity and emotional expression) noted: “{\it The agent expresses regret when making mistakes and expresses willingness to correct itself and it appears open to being helped.}”}

\rhl{Additionally, some participants (N = 28) explicitly cited empathy as their reason for helping the chatbot.} \mrhl{All of these participants came from the conditions with at least one anthropomorphism features.} \rhl{As P44 (from condition with human identity and non-verbal expression) stated: “\textit{I guess I felt empathy toward the AI. I felt annoyance toward whoever was not cleaning the workspace. I guess I would be happier if it functioned well.}” Similarly, P169 (from condition with emotional expression and non-verbal expression) shared: “\textit{There's a lot of sympathy generated in my interaction with the agent, so that encourages me to be helpful toward it when it encounters more trouble.}”}

\subsubsection{Reasons for Providing Emotional Support}
\rhl{Regarding participants’ willingness to provide emotional support and their underlying motivations, some participants from conditions with human identity (N = 35) expressed that they were inclined to comfort the chatbot because they imagined themselves in the agent’s position. Some responses referred to the agent using names or pronouns like “Jamie,” “him,” or “her.” For example, P97 (from the condition with human identity and emotional expression) said: “\textit{I am willing to comfort him because that is the right thing to do which will help him and also because I would like the same thing done for me should I be in the same situation as Jamie.}”}

\rhl{Participants from conditions with emotional expression (N = 38) reported that the chatbot’s display of emotion triggered sympathy, motivating them to provide emotional support. As P238 (from the condition with human identity and emotional expression) described: “\textit{I would feel sympathetic towards the agent if they were having a tough time and were feeling sad.}”}

\rhl{Some participants from the non-verbal expression only condition (N = 10) also stated that they comforted the chatbot because they perceived it as being sad. As P41 wrote: “\textit{I don't want it to be sad or express negative emotions, so I would try to comfort it and make it feel better. I know it can't really feel emotions, but it is clear to me that the agent knows getting labeling exercises wrong = bad.}” Similarly, P99 shared: “\textit{The sadness seems real and it seems cruel not to comfort it.}”}

\rhl{In the condition with all anthropomorphism features, a few participants (N = 9) mentioned a unique theme that did not emerge in other groups: they comforted the chatbot because it felt like a real human being. For instance, P2 stated: “\textit{It has a conscious similar to a real person.” Likewise, P37 said: “It felt at times like I was talking to a human being.}”}

\rhl{For those unwilling to provide emotional support, their main viewpoint was that there is no need to comfort an AI because it is just a program, not a human, and does not have genuine emotions. For example, P31 (from the condition without any anthropomorphism features) said: {\it It's not alive, it doesn't need to be comforted. If it was sad, it's because it was programmed to be sad; that isn't a genuine emotion.}}

%% file: sections/tables/interaction.tex
\begin{table}[t]
\caption{\rhl{Post-hoc comparison of the effect Identity*Emotional Expression*Non-verbal Expression on participants' providing emotional support. I: Identity, E: Emotional Expression, N: Non-verbal Expression. The significance levels are labeled ($p<0.05$: *, $p<0.01$: **, $p<0.001$: ***).}}
\label{tab:interaction-effect}
\resizebox{.55\textwidth}{!}{%
\arrayrulecolor{\rcolor}     
\color{\rcolor}              
\begin{tabular}{ccccl}
\cline{1-4}
\multicolumn{2}{c}{Condition (M, SD)}                      & t    & $p$                &  \\ \cline{1-4}
E, N (.50, .24)                     & w/o I, E, N (.24, .20) & 3.94 & .003** & \\
                                    \cline{1-4}
\multirow{6}{*}{I, E, N (.64, .24)} & w/o I, E, N (.24, .20) & 6.34 & \textless{}.001*** &  \\
                                    & I (.38, .23)         & 4.23 & \textless{}.001*** &  \\
                                    & E (.37, .22)         & 4.40  & \textless{}.001*** &  \\
                                    & N (.40, .27)         & 3.64 & .008**            &  \\
                                    & I, E (.43, .24)      & 3.67 & .007**            &  \\
                                    & I, N (.39, .26)      & 3.63 & .008**            &  \\ \cline{1-4}
\end{tabular}%
}
\end{table}
\arrayrulecolor{black}

%% file: sections/tables/regression.tex
\begin{table}[t]
\small
\caption{\rhl{The indirect mediating effect of empathy on the relationship between chatbot anthropomorphism features and participants' provision of emotional support and intention to provide instrumental support. SE = standard error; \(\beta\) = estimated effect; The significance levels are labeled ($p<0.05$: *, $p<0.01$: **, $p<0.001$: ***).}}
\label{path_coeff}
\resizebox{\textwidth}{!}{%
\arrayrulecolor{\rcolor} 
\color{\rcolor}
\begin{tabular}{cccccccc}
\hline
\multirow{2}{*}{Predictor} &
  \multirow{2}{*}{Mediator} &
  \multirow{2}{*}{Outcome} &
  \multirow{2}{*}{$\beta$} &
  \multirow{2}{*}{SE} &
  \multirow{2}{*}{$p$} &
  \multicolumn{2}{c}{95\% CI} \\ \cline{7-8} 
                                         &                                    &                      &      &      &                    & Lower & Upper \\ \hline
\multirow{4}{*}{Identity}                & \multirow{2}{*}{Cognitive Empathy} & Instrumental Support & .06 & .04 & .167               & -.03 & .15  \\
                                         &                                    & Emotional Support    & .11 & .07 & .117               & -.03 & .24  \\ \cline{2-8} 
                                         & \multirow{2}{*}{Affective Empathy} & Instrumental Support & .12 & .06 & .036*   & .01  & .23  \\
                                         &                                    & Emotional Support    & .18 & .08 & .026*  & .02  & .33  \\ \hline
\multirow{4}{*}{Emotional Expression}    & \multirow{2}{*}{Cognitive Empathy} & Instrumental Support & .13 & .06 & .032*              & .01  & .25  \\
                                         &                                    & Emotional Support    & .27 & .12 & .031*    & .03  & .51  \\ \cline{2-8} 
                                         & \multirow{2}{*}{Affective Empathy} & Instrumental Support & .26 & .11 & .012* & .06  & .47  \\
                                         &                                    & Emotional Support    & .40 & .15 & .006** & .11  & .69  \\ \hline
\multirow{4}{*}{Non-verbal   Expression} & \multirow{2}{*}{Cognitive Empathy} & Instrumental Support & .01 & .03 & .744               & -.05 & .07  \\
                                         &                                    & Emotional Support    & .02 & .05 & .742               & -.08 & .12   \\ \cline{2-8} 
                                         & \multirow{2}{*}{Affective Empathy} & Instrumental Support & .04 & .04 & .271               & -.03 & .11  \\
                                         &                                    & Emotional Support    & .06 & .06 & .264               & -.05 & .17  \\ \hline
\end{tabular}
}
\end{table}
\arrayrulecolor{black}

%% file: sections/05discussion.tex
\section{Discussion}
This study explores the effects of three anthropomorphism features of chatbots—identity, emotional expression, and non-verbal expression—on humans providing prosocial behaviors to chatbots, as well as the mediating effects of human cognitive empathy and affective empathy in this process. Through qualitative data, we further reveal the motivations behind people's prosocial behaviors toward chatbots. In the following sections, we discuss our findings, address the limitations of this study, and suggest directions for future research.

\subsection{The Effects of Chatbots' Anthropomorphism on Human Empathy and Prosocial Behaviors}

Our results indicate that the chatbot's human identity and emotional expression can enhance participants' cognitive empathy and affective empathy. 
This is consistent with previous research showing that anthropomorphism avatars and human-like language styles in customer service chatbots promote human empathy~\cite{janson2023leverage}. \rhl{Emotional expression had a larger effect size on both types of empathy compared to identity. This may be because identity relies on mechanisms such as emotional resonance to elicit empathy \cite{Riek2009How, cikara2011us}, whereas emotional expression can trigger empathetic responses more directly \cite{roberts1996empathy}.}

The chatbot's non-verbal expression did not have a significant effect on participants' empathy. This contrasts with previous studies that found emojis can elicit empathic neural responses in people~\cite{liao2021emoji}. One possible explanation is that the impact of emojis on human empathy is influenced by multiple factors, such as the selection of emojis, their frequency and number of use, and how different users interpret emojis~\cite{jeon2022differences, volkel2019understanding}, which warrants further exploration.

We also found that emotional expression of the chatbot enhanced participants' willingness to provide instrumental support to the chatbot. This result is consistent with previous findings that prosocial behavior between people can be promoted by emotional expressions~\cite{van2015emotional}. We further found that this promoting effect of emotional expression was fully mediated by participants' cognitive empathy and affective empathy. This highlights the important role of empathy as a prosocial emotion in the process of human prosocial behavior~\cite{hoffman2008empathy, Eisenberg1987}. 

In addition, we found that the direct effect of the identity of chatbots on participants' intention to provide instrumental support to the chatbot was not significant. However, mediation analysis showed that human identity could enhance participants' willingness to provide instrumental support by enhancing their affective empathy. This finding underscores the importance of psychological mechanisms in human-chatbot interactions. Specifically, it suggests that anthropomorphism alone might not be enough to elicit prosocial behavior. Rather, it is the empathy generated through anthropomorphism that drives such behaviors.
\rhl{Notably, only affective empathy mediated the effect of identity, whereas cognitive empathy did not. This may be because the chatbot’s human-like identity serves as a social cue that increases users’ perceived self–other overlap \cite{Herrera2018Building}, thereby eliciting more in-group affective resonance in this context \cite{cikara2011us, tajfel1979integrative}.
While the cognitive empathy pathway involves cognitive modeling of the other’s situation, which relies more heavily on explicit descriptions of the context through language \cite{Decety2006Human}. Thus, the identity cue alone may not be sufficient to activate this more deliberate form of empathy. Despite this, our qualitative findings revealed that the human-like identity of the chatbot made it easier for some participants to engage in perspective-taking, which in turn increased their willingness to provide emotional support. This may be because these participants have a higher tendency for cognitive empathy. Future research could further explore how users' traits influence their willingness to provide prosocial behavior toward chatbots.
}

Non-verbal expression has neither a direct nor indirect effect on promoting people's willingness to provide instrumental support to the chatbot. This differs from previous research showing that emojis can promote prosocial behavior in online interactions between human beings \cite{zhang2021influence}. However, we also find that non-verbal cues in chatbots lead to an increase in participants' emotional support provided to chatbots. A possible explanation is that, non-verbal expression is a more fundamental stimulus that can activate human unconscious social scripts related to prosocial behavior~\cite{nass_CASA,nass_mindlessness}. Specifically, during the conversation with chatbots, participants might have unconsciously provided more emotional support, but when asked in the post-survey about their willingness to offer instrumental help, they might have shifted to more conscious, deliberate thinking, which did not increase their likelihood of offering help.
\rhl{Similarly, the finding that the effect of non-verbal expression on emotional support was not mediated by either cognitive or affective empathy further supports this interpretation. Prior research has shown that people can rapidly and subconsciously interpret non-verbal cues, bypassing cognitive processing \cite{ambady1992thin, de2006towards}. This suggests that the impact of non-verbal expression on emotional support may primarily involve subconscious mechanisms. Considering our qualitative results, another possibility is that the non-verbal expressions activated other psychological pathways, such as the perception of chatbot emotions, that influenced users’ prosocial behaviors independently of empathy \cite{wright2012emotional}. Future work may explore these alternative mechanisms to better understand how non-verbal cues can shape human prosocial behaviors toward chatbots.}

\rhl{We also observed a potential interaction effect among the three anthropomorphism features in facilitating emotional support. When all three features were present, participants provided significantly higher levels of emotional support. This suggests that the relationship between prosocial behavior and anthropomorphism features is not linear. Rather, once the number of \mrhl{anthropomorphism features} surpasses a certain threshold, users become more willing to engage in prosocial behaviors. Findings from our qualitative analysis further support this interpretation: only when all three features were present did some participants report comforting the chatbot because it felt particularly human-like.}

\subsection{Human Understanding of Prosocial Behaviors toward Chatbots}

Our qualitative results indicate that participants' motivations for providing instrumental support and emotional support are: (1) empathy for the chatbot’s situation, (2) perceiving the chatbot as human-like.

These motivations support our findings on how chatbot anthropomorphism features and human empathy promote people's prosocial behaviors toward chatbots. 
\rhl{A human-like identity led participants to put themselves in the chatbot's shoes and refer to the agent in more personalized ways. 
It supports previous work suggesting that human-like identity cues can promote social categorization in human–AI interaction \cite{janson2023leverage, COWELL2005281}. 
By shaping how users refer to and relate to the chatbot, identity cues may activate group-based emotional resonance similar to mechanisms observed in human–human interactions.
Emotional expression allowed users to perceive the chatbot’s affective state, thereby triggering empathetic responses. 
This aligns with previous findings in both human–human and human–AI interaction, where emotional expression has been shown to reliably predict empathic responses by making emotional states more accessible and salient \cite{janson2023leverage, van2015emotional, roberts1996empathy}.
Notably, even in the presence of only non-verbal expression, some users still perceived the chatbot as experiencing “real sadness,” suggesting that non-verbal cues can have a strong impact on users’ perception of the chatbot’s emotional state.
This extends prior research showing that non-verbal cues can influence affective judgments, even in the absence of verbal language \cite{thompson2014sex, derks2008emoticons}. 
}

Participants who expressed negative attitudes toward providing emotional support stated that they did not want to comfort a non-human entity without genuine emotions or with only simulated emotions. 
\rhl{This finding differs from previous research showing that users express compassion toward chatbots based on emotional empathy \cite{Lee2019caring}. 
One reason for this discrepancy may be that many of the participants who expressed this view in our study were assigned to conditions without emotional expression, whereas in Lee et al.~\cite{Lee2019caring}, the chatbot consistently displayed emotional expression. However, we also observed that some participants in our emotional expression conditions still resisted providing emotional support.
One possible explanation is individual cognitive differences: some people are more inclined to form emotional connections with non-human agents, while others remain skeptical of AI-generated emotions, believing that emotional support should be reserved for humans with genuine feelings. Another contributing factor may be the short-term, task-oriented context of our study, in which the chatbot was perceived more as a tool than a relational partner. In contrast, prior work was conducted in an emotionally driven setting \cite{Lee2019caring}. In contexts that emphasize the human–chatbot relationship, such as companion bots, users may feel more comfortable expressing emotional support.}

\subsection{Design Implications}

This study has several design implications. 
In human-AI collaborations, there are many scenarios where humans and AI can complement each other’s strengths and AI require human help, such as in human-AI decision-making \cite{li2024overconfident, zhang2020effect}. \rhl{In these cases, designers can enhance the chatbot’s human-like identity (e.g., a human profile photo, a human name of the AI assistant) and add appropriate emotional expressions (e.g., expressing nervous or sorry when making mistakes) to encourage users to provide task-related assistance. 
Given that some users may still view the chatbot as a mere tool, designers may pay additional attention to the identity setting. Designers could adopt framing strategies that emphasize the chatbot as a collaborative partner rather than a passive system. For example, the chatbot can use first-person language that emphasizes interdependence (e.g., “Let’s work on this together” or “I could use your input here”), or display expressions of vulnerability (e.g., uncertainty) \cite{li2025confidence} to signal the need for user support. These strategies may help shift user expectations and encourage more reciprocal behaviors in human–AI teamwork.}

Our findings can also be applied to scenarios where eliciting prosocial behavior toward chatbots can bring positive effects to humans. For example, existing research shows that providing emotional support can also improve the emotional well-being of the provider \cite{NIELSEN2022260, zhu2025benefits}. Based on our results, designers can create chatbots with these anthropomorphism features to encourage human emotional support behaviors toward chatbots to improve human emotional health. For example, a companion chatbot could express emotions such as sadness, encouraging users to offer comfort and allowing them to experience satisfaction from providing emotional support. 
\rhl{Designers may also consider incorporating emotionally expressive emojis to activate users’ automatic social responses and foster prosocial behavior toward chatbots during interaction. These perspectives extend CSCW’s ongoing exploration of how socio-emotional dynamics in human–AI interactions can be designed to support not only task collaboration, but also human well-being.}

\rhl{The potential interaction effects among different anthropomorphism features also warrant careful consideration by future designers and researchers. Our findings suggest that combining multiple anthropomorphism features can produce \mrhl{an interaction} effect where the overall impact on user perception and behavior exceeds the sum of individual features. In other words, there may be a “tipping point” beyond which the influence of anthropomorphism on users increases rapidly. This provides an important extension to prior work on chatbot anthropomorphism \cite{janson2023leverage, seeger2018designing}. On the one hand, this presents an opportunity: by strategically combining anthropomorphism features, designers may achieve greater user engagement than through isolated features alone. \mrhl{On the other hand, in scenarios where the degree of anthropomorphism must be carefully managed, such as when emphasizing the instrumental nature of chatbots to avoid overtrust and over-reliance \cite{waytz2014mind}, these interaction effects may pose a challenge.} One feasible strategy is to adaptively adjust the use of anthropomorphism features. For example, minimizing elements such as emotional expression during the collaboration process, but introducing them afterward when the AI requires help or feedback.
In summary, designers need to anticipate that simply layering \mrhl{anthropomorphism features} can lead to non-linear increases in certain perceptions (e.g., affective empathy) of users, potentially complicating efforts to maintain appropriate human–AI boundaries.}

Moreover, our results in relation to empathy can also be applied to scenarios where eliciting empathy toward chatbots is necessary, such as conflict rehearsal with chatbots \cite{shaikh2024rehearsal}, and reducing verbal bullying and chatbot abuse \cite{keijsers2021s}. Based on our findings, designers can consider enhancing chatbot human identity and emotional expression to foster user empathy. For instance, in conflict rehearsal scenarios, designing a chatbot with a human-like identity and the ability to appropriately express emotions may better elicit user empathy and help users engage more fully in the rehearsal context. 
\rhl{These application scenarios should also take into account the differences in how cognitive and affective empathy function across different contexts. In goal-oriented interactions, where mutual understanding and coordination are key, features that support cognitive empathy may be more effective, such as clear identity cues which can enhance users' perspective taking or a detailed description of the chatbot situation. In contrast, emotionally supportive tasks can benefit more from features that evoke affective empathy, such as expressive language or emotional displays. Designing with empathy types in mind can help create more adaptive and context-sensitive chatbot experiences.}

\subsection{Limitations and Future Work}
This study has several limitations. First, we only examined human prosocial behaviors toward chatbots in a collaborative task where the chatbot made mistakes. However, prosocial behaviors toward chatbots can occur in other contexts, such as offering help outside of a collaborative scenario. Future studies could explore human prosocial behaviors toward chatbots in different contexts.

Meanwhile, for instrumental support, we only measured participants’ intentions rather than their actual behaviors. Although intentions are often an indicator of behavior, there are still some differences between them \cite{webb2006does}. We suggest that future research could measure actual behaviors of providing instrumental support, such as whether people actively help chatbots correct errors or provide feedback on their mistakes.

\rhl{Second, the study design has certain limitations. The manipulation of anthropomorphism could be further expanded. In our study, we used a back-view human profile photo to minimize possible confounders. However, this conservative design choice may have constrained the strength of the identity manipulation. 
Similarly, future experiments could use a simple logo as the profile photo for conditions without human identity instead of a robot avatar to further strengthen the manipulation of identity in terms of anthropomorphism.
Furthermore, AI's explanation for its errors focused primarily on hardware-related issues, which may have limited the opportunities and willingness of participants to offer instrumental support during the interaction. Future work may consider some direct reasons such as the quality of training data, which make it easier to include an opportunity for the user to provide real instrumental support.} 

In addition, this study examined human prosocial behaviors toward chatbots in a short-term experiment. Future research could consider conducting long-term studies to explore how chatbots can elicit prosocial behavior over long periods and encourage the development of lasting prosocial tendencies. This approach would not only help researchers better understand human prosociality but could also yield greater benefits and promote social good.

\mrhl{Moreover, when researchers and designers use anthropomorphism features to enhance users' empathy toward AI and willingness to cooperate, they should also recognize and consider the potential risks. Highly anthropomorphized AI chatbots can make users reliant too much on them~\cite{akbulut2024all}, which can reduce human agency and negatively affect the outcomes of human-AI collaboration. In the long term, anthropomorphism can also cause users to form emotional bonds with human-like AI, leading to excessive self-disclosure and risks to user privacy~\cite{akbulut2024all}. Therefore, while promoting empathy and prosocial behavior can benefit well-being and human-AI collaboration, the associated risks of high anthropomorphism must be considered and addressed in future design and research.}

\rhl{Finally, this study only investigated the effects of basic anthropomorphism features. Future work could investigate how these foundational anthropomorphism features interact with the advanced linguistic and social capabilities of LLM-driven chatbots. This would provide a more comprehensive understanding of anthropomorphism as it manifests in contemporary AI systems.}

%% file: sections/06conclusion.tex
\section{Conclusion}
With the development and widespread use of chatbots, there is growing interest in leveraging chatbots to elicit human prosocial behaviors toward them due to the wide range of benefits this can offer. However, the factors that motivate human prosocial behavior toward chatbots remain unclear. Addressing this gap, our study finds that anthropomorphic features of chatbots can encourage human prosocial behaviors toward them, with human empathy mediating the effects of specific anthropomorphic features including human-like identity and emotional expression. Our qualitative analysis further reveals that the desire for positive human-chatbot collaborative outcomes and individuals’ own prosocial tendencies are motivators for engaging in prosocial behaviors toward chatbots. Overall, these findings provide design implications to encourage human prosocial behavior toward chatbots in human-chatbot interactions, which could benefit individuals, human-chatbot collaboration, and broader social good.

%% file: appendix/descriptive-results-for-each-group.tex
\clearpage
\section{Descriptives and ANCOVA Result Details}
\label{sec: Descriptives}
 
\begin{table}[H]
\caption{\rhl{Descriptives for Sec.~\ref{sec:manipulation-check} (Manipulation Check)}}
\arrayrulecolor{\rcolor}     
\color{\rcolor}  
\resizebox{\textwidth}{!}{
\begin{tabular}{|cc|cccccccc|}
\multicolumn{2}{|c|}{Group \#}                                  & 1       & 2       & 3       & 4       & 5       & 6     & 7       & 8     \\
\multicolumn{2}{|c|}{Identity}                                  & Chatbot & Chatbot & Chatbot & Chatbot & Human   & Human & Human   & Human \\
\multicolumn{2}{|c|}{Emotional   Expression}                    & Low     & Low     & High    & High    & Low     & Low   & High    & High  \\
\multicolumn{2}{|c|}{Non-verbal   Expression}                   & Without & With    & Without & With    & Without & With  & Without & With  \\
\multicolumn{1}{|c|}{\multirow{2}{*}{Perceived   Anthropomorphism}} & Mean & 1.98 & 2.67 & 3.64 & 4.49 & 2.97 & 3.18 & 4.17 & 4.25 \\
\multicolumn{1}{|c|}{}                                   & SD   & 1.04    & 1.18    & 1.22    & 1.28    & 1.02    & 1.11  & 1.37    & 1.33  
\end{tabular}
}
\end{table}
\arrayrulecolor{black}    

\begin{table}[H]
\caption{\rhl{Descriptives for Sec.~\ref{sec:H1}, and Sec.~\ref{sec:H2H3} (H1, H2 and H3)}}
\resizebox{\textwidth}{!}{
\begin{tabular}{|cc|cccccccc|}
\multicolumn{2}{|c|}{Group \#}                                  & 1       & 2       & 3       & 4       & 5       & 6     & 7       & 8     \\
\multicolumn{2}{|c|}{Identity}                                  & Chatbot & Chatbot & Chatbot & Chatbot & Human   & Human & Human   & Human \\
\multicolumn{2}{|c|}{Emotional   Expression}                    & Low     & Low     & High    & High    & Low     & Low   & High    & High  \\
\multicolumn{2}{|c|}{Non-verbal   Expression}                   & Without & With    & Without & With    & Without & With  & Without & With  \\
\multicolumn{1}{|c|}{\multirow{2}{*}{Instrumental Support}}         & Mean & 3.70 & 3.83 & 3.84 & 3.90 & 3.68 & 3.83 & 4.45 & 4.27 \\
\multicolumn{1}{|c|}{}                                   & SD   & 1.06    & 1.12    & 1.13    & 0.94    & 1.14    & 1.09  & 0.68    & 0.98  \\
\multicolumn{1}{|c|}{\multirow{2}{*}{Emotional Support}} & Mean & 0.24    & 0.40    & 0.37    & 0.50    & 0.38    & 0.39  & 0.43    & 0.64  \\
\multicolumn{1}{|c|}{}                                   & SD   & 0.20    & 0.27    & 0.22    & 0.24    & 0.23    & 0.26  & 0.24    & 0.24  \\
\multicolumn{1}{|c|}{\multirow{2}{*}{Cognitive Empathy}} & Mean & 2.67    & 3.02    & 3.42    & 3.46    & 3.15    & 3.00  & 3.72    & 3.70  \\
\multicolumn{1}{|c|}{}                                   & SD   & 1.00    & 1.13    & 0.94    & 0.95    & 0.98    & 1.00  & 0.71    & 0.92  \\
\multicolumn{1}{|c|}{\multirow{2}{*}{Affective Empathy}} & Mean & 2.52    & 3.22    & 4.48    & 4.57    & 3.65    & 3.65  & 4.92    & 5.13  \\
\multicolumn{1}{|c|}{}                                   & SD   & 1.51    & 1.71    & 1.53    & 1.49    & 1.69    & 1.50  & 1.43    & 1.48 
\end{tabular}
}
\end{table}

\begin{table}[H]
\caption{Number of participants in each experimental group}
\label{table:number of participants}
\begin{tabular}{cccc}
\hline
Identity & Emotional Expression & Non-verbal Expression & Number of Participants \\ \hline
Chatbot  & Low                  & Without               & 30                     \\
Chatbot  & Low                  & With                  & 30                     \\
Chatbot  & High                 & Without               & 31                     \\
Chatbot  & High                 & With                  & 31                     \\
Human    & Low                  & Without               & 31                     \\
Human    & Low                  & With                  & 30                     \\
Human    & High                 & Without               & 31                     \\
Human    & High                 & With                  & 30                     \\ \hline
\end{tabular}
\end{table}

\begin{table}[H]
\caption{\rhl{ANCOVA \mrhl{(Type III sum of squares)} result details for the effects of anthropomorphism features on participants' providing emotional support. The significance levels are labeled ($p<0.05$: *, $p<0.01$: **, $p<0.001$: ***).}}
\label{tab:ANCOVA Emotional Support}
\arrayrulecolor{\rcolor}     
\color{\rcolor}      
\begin{tabular}{cccc}
\hline
Variable                                              & F                & $p$                 & $\eta^2$         \\ \hline
Identity                                              & 6.86             & 0.009**             & 0.024            \\
Emotional Expression                                  & 13.82            & \textless{}0.001*** & 0.048            \\
Non-verbal Expression                                 & 20.47            & \textless{}0.001*** & 0.071            \\
Identity*Emotional Expression                         & 0.20             & 0.657               & 0.001            \\
Identity*Non-verbal   Expression                      & 0.08             & 0.781               & \textless{}0.001 \\
Emotional   Expression*Non-verbal Expression          & 1.50             & 0.222               & 0.005            \\
Identity*Emotional   Expression*Non-verbal Expression & 3.68             & 0.056               & 0.013            \\
Age                                                   & \textless{}0.001 & 0.978               & \textless{}0.001 \\
Gender                                                & 7.42             & 0.007**             & 0.026            \\
Education                                             & 0.13             & 0.724               & \textless{}0.001 \\
General Attitude toward AI                            & 3.11             & 0.079               & 0.013            \\ \hline
\end{tabular}
\end{table}

\begin{table}[H]
\caption{\rhl{ANCOVA \mrhl{(Type III sum of squares)} result details for the effects of anthropomorphism features on participants' intention to provide instrumental support. The significance levels are labeled ($p<0.05$: *, $p<0.01$: **, $p<0.001$: ***).}}
\label{tab:ANCOVA Instrumental Support}
\color{\rcolor}     
\begin{tabular}{cccc}
\hline
Variable                                              & F     & $p$                 & $\eta^2$         \\ \hline
Identity                                              & 2.01  & 0.158               & 0.007            \\
Emotional Expression                                  & 4.14  & 0.037*              & 0.016            \\
Non-verbal Expression                                 & 0.60  & 0.441               & 0.002            \\
Identity*Emotional Expression                         & 2.91  & 0.089               & 0.011            \\
Identity*Non-verbal   Expression                      & 0.01  & 0.946               & \textless{}0.001 \\
Emotional   Expression*Non-verbal Expression          & 0.91  & 0.340               & 0.003            \\
Identity*Emotional   Expression*Non-verbal Expression & 0.26  & 0.661               & 0.001            \\
Age                                                   & 0.59  & 0.443               & 0.002            \\
Gender                                                & 2.02  & 0.157               & 0.008            \\
Education                                             & 2.05  & 0.153               & 0.008            \\
General Attitude toward AI                            & 20.53 & \textless{}0.001*** & 0.077            \\ \hline
\end{tabular}
\end{table}

\begin{table}[H]
\caption{\rhl{ANCOVA \mrhl{(Type III sum of squares)} result details for the effects of anthropomorphism features on participants' cognitive empathy. The significance levels are labeled ($p<0.05$: *, $p<0.01$: **, $p<0.001$: ***).}}
\label{tab:ANCOVA Cognitive Empathy}
\color{\rcolor}     
\begin{tabular}{cccc}
\hline
Variable                                              & F                & $p$                 & $\eta^2$         \\ \hline
Identity                                              & 4.17             & 0.042*              & 0.015            \\
Emotional Expression                                  & 19.64            & \textless{}0.001*** & 0.072            \\
Non-verbal Expression                                 & 0.57             & 0.453               & 0.002            \\
Identity*Emotional Expression                         & \textless{}0.001 & 0.998               & \textless{}0.001 \\
Identity*Non-verbal   Expression                      & 1.01             & 0.316               & 0.004            \\
Emotional   Expression*Non-verbal Expression          & 0.21             & 0.650               & 0.001            \\
Identity*Emotional   Expression*Non-verbal Expression & 0.49             & 0.484               & 0.002            \\
Age                                                   & 4.92             & 0.027*              & 0.018            \\
Gender                                                & 7.87             & 0.005**             & 0.029            \\
Education                                             & 0.18             & 0.669               & \textless{}0.001 \\
General Attitude toward AI                            & 1.46             & 0.228               & 0.005            \\ \hline
\end{tabular}
\end{table}

\begin{table}[H]
\caption{\rhl{ANCOVA \mrhl{(Type III sum of squares)} result details for the effects of anthropomorphism features on participants' affective empathy. The significance levels are labeled ($p<0.05$: *, $p<0.01$: **, $p<0.001$: ***).}}
\label{tab:ANCOVA Affective Empathy}
\color{\rcolor}     
\begin{tabular}{cccc}
\hline
Variable                                              & F     & $p$                 & $\eta^2$         \\ \hline
Identity                                              & 8.63  & 0.004**             & 0.027            \\
Emotional Expression                                  & 49.01 & \textless{}0.001*** & 0.154            \\
Non-verbal Expression                                 & 3.14  & 0.078               & 0.010            \\
Identity*Emotional Expression                         & 0.95  & 0.331               & 0.003            \\
Identity*Non-verbal   Expression                      & 0.14  & 0.706               & \textless{}0.001 \\
Emotional   Expression*Non-verbal Expression          & 0.33  & 0.564               & 0.001            \\
Identity*Emotional   Expression*Non-verbal Expression & 0.88  & 0.350               & 0.003            \\
Age                                                   & 1.80  & 0.183               & 0.006            \\
Gender                                                & 4.92  & 0.028*              & 0.015            \\
Education                                             & 0.03  & 0.863               & \textless{}0.001 \\
General Attitude toward AI                            & 17.00 & \textless{}0.001*** & 0.053            \\ \hline
\end{tabular}
\end{table}

\arrayrulecolor{black}  

%% file: appendix/scale.tex
\section{Scales}
\label{appendix:scales}

\subsection{General Attitude towards AI Scale}
General Attitude towards AI Scale is a 5-points Likert Scale with 5 items \cite{SCHEPMAN2020100014}, including (* marks reversed items.):
\begin{itemize}
    \item I am interested in using artificially intelligent systems in my daily life.
    \item There are many beneficial applications of Artificial Intelligence.
    \item Artificial Intelligence is exciting.
    \item I think Artificial Intelligence is dangerous.*
    \item Organizations use Artificial Intelligence unethically.*
\end{itemize}

\subsection{Perceived Anthropomorphism Scale}
General Attitude towards AI Scale is a 7-points Likert Scale with 7 items \cite{seeger2018designing}, including:
\begin{itemize}
    \item The agent has a mind of its own.
    \item The agent has intentions.
    \item The agent has free will.
    \item The agent has consciousness.
    \item The agent has desires.
    \item The agent has beliefs.
    \item The agent has the ability to experience emotions.
\end{itemize}

\subsection{Empathy Scale}
Affective Empathy Scale is a 7-points Likert Scale with 6 items \cite{Baston1997perspective, Herrera2018Building}. Participants were asked to answer "{\it To what extend did you experience the following emotions while talking with the agent:}"
\begin{itemize}
    \item Sympathetic
    \item Softhearted
    \item Warm
    \item Compassionate
    \item Tender
    \item Moved
\end{itemize}

Cognitive Empathy Scale is a 5-points Likert Scale with 3 items \cite{davis1983measuring}, including (* marks reversed items.):
\begin{itemize}
    \item I believe the agent’s unsatisfactory performance came from the poor working conditions.
    \item In the previous conversation, I found it difficult for me to look at the problems facing the agent from the agent‘s point of view.*
    \item In the previous conversation, I tried to understand the agent's difficulties by imagining the agent's situation.
\end{itemize}